\documentclass[manuscript,screen]{acmart}

\AtBeginDocument{%
  }



\setcopyright{acmlicensed}
\copyrightyear{2026}
\acmYear{2026}
\acmDOI{XXXXXXX.XXXXXXX}

\acmJournal{TOSEM}
\acmVolume{1}
\acmNumber{1}
\acmArticle{1}
\acmMonth{8}

\usepackage{xspace}
\usepackage{flushend}
\usepackage{enumitem}
\usepackage[ruled,vlined,linesnumbered]{algorithm2e}
\usepackage{multicol}
\usepackage{xcolor}
\usepackage[skins]{tcolorbox}
\usepackage{placeins}
\usepackage{amsmath}
\usepackage{bm}
\usepackage{subcaption}
\usepackage{amsthm}
\usepackage{listings}
\theoremstyle{acmdefinition}
\newtheorem{exmp}{Example}[section]
\usepackage{multirow}
\usepackage{booktabs}
\usepackage{tabularx}
\usepackage{pifont}
\usepackage[normalem]{ulem}
\usepackage{color, colortbl}
\usepackage{makecell}
\usepackage{dashbox}
\definecolor{mygreen}{RGB}{54,120,33}  
\definecolor{myred}{RGB}{192,0,0}    
\definecolor{graybg}{RGB}{230,230,230}  
\definecolor{lightgreen}{RGB}{200,255,200}  
\definecolor{darkgreen}{RGB}{144,238,144}  
\definecolor{lightblue}{RGB}{36,100,171}  
\definecolor{lightred}{RGB}{177,24,45}  

\newcommand{\todo}[1]{\textcolor{black}{#1}}

\newcommand{\cyh}[1]{\textcolor{black}{#1}}
\newcommand{\revision}[1]{\textcolor{black}{#1}}
\newcommand{\camera}[1]{\textcolor{black}{#1}}
\newcommand{\tool}{\textsc{VulWeaver}\xspace}

\setlist{nosep, leftmargin=*}  

\begin{document}

\title{\tool: Weaving Broken Semantics for Grounded Vulnerability Detection}

\author{Yiheng Cao}
\authornote{Y. Cao, Y. Chen, X. Hu, B. Chen, J. Deng, Z. Zhou, S. Wu, Y. Huang, X. Du, X. Peng are with the College of Computer Science and Artificial Intelligence.}
\orcid{0009-0007-6101-8270}
\affiliation{
\department{College of Computer Science and Artificial Intelligence}
\institution{Fudan University}
\city{Shanghai}
\country{China}
}

\author{Yihao Chen}
\authornotemark[1]
\orcid{0009-0001-2023-8812}
\affiliation{
\department{College of Computer Science and Artificial Intelligence}
\institution{Fudan University}
\city{Shanghai}
\country{China}
}

\author{Xin Hu}
\authornotemark[1]
\orcid{0009-0009-2794-9623}
\affiliation{
\department{College of Computer Science and Artificial Intelligence}
\institution{Fudan University}
\city{Shanghai}
\country{China}
}

\author{Bihuan Chen}
\authornotemark[1]
\orcid{0000-0001-7238-7492}
\affiliation{
\department{College of Computer Science and Artificial Intelligence}
\institution{Fudan University}
\city{Shanghai}
\country{China}
}

\author{Jiayi Deng}
\authornotemark[1]
\orcid{0009-0000-7935-3578}
\affiliation{
\department{College of Computer Science and Artificial Intelligence}
\institution{Fudan University}
\city{Shanghai}
\country{China}
}

\author{Zhuotong Zhou}
\authornotemark[1]
\orcid{0009-0006-7819-9656}
\affiliation{
\department{College of Computer Science and Artificial Intelligence}
\institution{Fudan University}
\city{Shanghai}
\country{China}
}

\author{Susheng Wu}
\authornotemark[1]
\orcid{0009-0005-2169-7032}
\affiliation{
\department{College of Computer Science and Artificial Intelligence}
\institution{Fudan University}
\city{Shanghai}
\country{China}
}

\author{Yiheng Huang}
\authornotemark[1]
\orcid{0009-0003-4722-3658}
\affiliation{
\department{College of Computer Science and Artificial Intelligence}
\institution{Fudan University}
\city{Shanghai}
\country{China}
}

\author{Xueying Du}
\authornotemark[1]
\orcid{0009-0005-0004-9183}
\affiliation{
\department{College of Computer Science and Artificial Intelligence}
\institution{Fudan University}
\city{Shanghai}
\country{China}
}

\author{Xingman Chen}
\authornote{X. Chen, M. Li are with Huawei Technologies Co., Ltd.}
\orcid{0000-0002-7120-9261}
\affiliation{
\institution{Huawei Technologies Co., Ltd}
\city{Dongguan}
\country{China}
}

\author{Miaohua Li}
\authornotemark[2]
\orcid{0009-0006-9681-8342}
\affiliation{
\institution{Huawei Technologies Co., Ltd}
\city{Dongguan}
\country{China}
}

\author{Xin Peng}
\authornotemark[1]
\authornote{X. Peng is the corresponding author.}
\orcid{0000-0003-3376-2581}
\affiliation{
\department{College of Computer Science and Artificial Intelligence}
\institution{Fudan University}
\city{Shanghai}
\country{China}
}
\renewcommand{\shortauthors}{Cao et al.}
\newcommand{\basecell}[1]{\cellcolor{red!6}{#1}} 

\newcommand{\adcell}[2]{\cellcolor{red!#1}{#2}}  
\definecolor{level0}{HTML}{FFF5F5} 
\definecolor{level1}{HTML}{FFD9D9} 
\definecolor{level2}{HTML}{FF9999} 
\definecolor{level3}{HTML}{E04040} 
\definecolor{level4}{HTML}{8B0000}


\begin{abstract}
Detecting vulnerabilities in source code remains critical yet challenging, as conventional static analysis tools construct inaccurate program representations, while existing LLM-based approaches often miss essential vulnerability context and lack grounded reasoning.
In this paper, we introduce \tool, a novel LLM-based approach that weaves broken program semantics into accurate representations and extracts holistic vulnerability context for grounded vulnerability detection.
\tool first constructs an enhanced unified dependency graph (UDG) by integrating deterministic rules with LLM-based semantic inference to address static analysis inaccuracies.
It then extracts holistic vulnerability context by combining explicit contexts from program slicing with implicit contexts, including usage, definition, and declaration information.
Finally, \tool employs meta-prompting with vulnerability type specific expert guidelines to steer LLMs through systematic reasoning, aggregated via majority voting for robustness.
Extensive experiments on \textsc{PrimeVul4J} dataset show that \tool achieves a precision of \todo{\revision{0.82}}, recall of \todo{\revision{0.71}}, and F1-score of \todo{\revision{0.76}}, outperforming state-of-the-art learning-based, LLM-based, and agent-based baselines by \todo{\revision{25}}\%, \todo{\revision{17}}\%, and \todo{\revision{21}}\% in F1-score, respectively.
Notably, \tool attains a VP-S score of \todo{\revision{0.58}}, \todo{\revision{164}}\% higher than the best baseline, confirming its strong discriminative power in distinguishing vulnerable code from patched counterparts.
\tool also demonstrates cross-language generalizability on the C/C++ \textsc{PrimeVul} dataset with minimal adaptation, achieving an F1-score of \todo{\revision{0.78}}.
For practical usefulness, \tool detected \todo{26} true vulnerabilities across \todo{9} real-world Java projects, with \todo{15} confirmed by developers and \todo{5} CVE identifiers assigned.
In industrial deployment, \tool identified \todo{40} confirmed vulnerabilities in an internal repository.
\end{abstract}


\maketitle


\section{Introduction}\label{sec:intro}
Security vulnerabilities are a major threat to software systems, leading to incidents such as crashes and data breaches. Therefore, early detection during development is paramount for ensuring software security~\cite{lekssays2025llmxcpg}. 
However, effective detection is increasingly challenged by the sheer scale and complexity of modern software ecosystems. In such large-scale scenarios, complex dependencies and heterogeneous build environments often hinder build automation. This renders dynamic analysis and binary-level static analysis infeasible, thereby restricting detection approaches to pure source code analysis~\cite{yu2025cxxcrafter}.
Unfortunately, traditional source code vulnerability detection approaches, such as static application security testing (SAST), depend on rigid rules and pattern matching. Although they offer broad coverage, they lack deep semantic understanding of code, leading to high false positive rates and burdening developers with excessive manual verification~\cite{widyasari2025let}.

To address the rigidity of traditional tools and capture complex code semantics, researchers have explored learning-based approaches, evolving from sequence models~\cite{li2016vulpecker,li2018vuldeepecker,li2021sysevr} to graph-based models that better represent code structure and dependencies~\cite{zhou2019devign,cheng2021deepwukong,steenhoek2024dataflow}.
More recently, large language models (LLMs) have advanced vulnerability detection via strategies such as chain-of-thought (CoT) prompting~\cite{nong2024chain, steenhoek2024comprehensive, tamberg2025harnessing, ullah2024llms, zhang2024prompt, zhou2024large}, in-context learning~\cite{lu2024grace}, and fine-tuning~\cite{du2024generalization, yang2024security}.
However, most of these approaches are limited to single-function contexts, missing inter-procedural information necessary for detecting complex vulnerabilities.

\revision{\textbf{Limitations.} To bridge this gap, recent studies incorporate source code level static analysis tools~\cite{joern, codeql} to extract inter-procedural context~\cite{lekssays2025llmxcpg,lu2024grace}.}
Typically, these approaches construct inter-procedural program representations (e.g., ICFG) to serve as the foundation for extracting code context via program slicing, which is then analyzed by LLM with fine-tuning~\cite{lekssays2025llmxcpg} or CoT prompting~\cite{lu2024grace}.
\revision{However, significant limitations persist regarding the completeness of the vulnerability context provided to LLMs and the quality of LLM reasoning over that context. In this work, we refer to the code context of the potential vulnerable part as \emph{vulnerability context}.}

\revision{\textit{\textbf{Limitation 1: The vulnerability context provided to LLMs is incomplete.}}}
\revision{Accurate vulnerability detection requires the LLM to analyze complete vulnerability context surrounding a suspicious code location. However, existing approaches fail to provide such context for two compounding reasons.}
\revision{First, the program representations that serve as the foundation for context extraction are themselves inaccurate. Scalable static analysis tools rely on conservative approximations to handle complex language features (e.g., polymorphism, and reflection), which inevitably leads to either over-approximation (e.g., adding spurious edges for polymorphic calls) or under-approximation (e.g., omitting edges for reflective dispatch). These inaccuracies propagate directly into the extracted context, introducing missing or misleading information.}
\revision{Second, even given a perfectly accurate program representation, existing context extraction strategies are insufficient. Most approaches restrict analysis to single-function boundaries~\cite{ding2024vulnerability, widyasari2025let, zhu2025specification}, and even those that attempt inter-procedural slicing typically follow only unidirectional paths along explicit dependencies, omitting semantically critical but structurally disjoint contexts such as global configurations~\cite{lekssays2025llmxcpg,lu2024grace}. As a result, LLMs receive fragmented vulnerability contexts that lack the completeness needed for reliable detection.}

\revision{\textit{\textbf{Limitation 2: LLM reasoning over vulnerability context is unstructured and ungrounded.}}}
\revision{Even with complete context, LLMs frequently misclassify vulnerabilities because they rely on superficial pattern matching rather than explicit logical reasoning~\cite{ding2024vulnerability}. Unlike security experts who follow systematic, step-by-step deduction tailored to each vulnerability type, current LLM-based approaches lack structured reasoning guidance. They often capture spurious correlations, over-relying on lexical cues (e.g., sensitive variable names) while overlooking critical control- and data-flow semantics. Hence, their decisions remain ungrounded, especially when distinguishing nuanced differences between vulnerable and non-vulnerable code.}

\revision{\textbf{Our Approach.} We introduce \tool, a novel approach driven by a single guiding principle that \emph{accurate vulnerability detection requires both complete vulnerability context and structured, vulnerability-aware reasoning over that context.} All design choices in \tool follow from this principle.}
\revision{To address \textbf{Limitation 1}, \tool first constructs an enhanced unified dependency graph (UDG) by unifying control flow, data dependency, and call graphs from a static analysis tool, and enhancing them via a neuro-symbolic strategy that integrates deterministic syntax-directed constraints with LLM-based semantic inference. This repairs precisely those semantic relationships that are critical for vulnerability context extraction yet beyond the reach of scalable static analysis tools. Building on this enhanced representation, \tool then extracts a holistic vulnerability context by integrating both explicit and implicit contexts, capturing complete information without introducing excessive noise.}
\revision{To address \textbf{Limitation 2}, \tool employs a meta-prompting framework~\cite{zhang2023meta, suzgun2024meta} that steers LLM reasoning using expert-defined, vulnerability type specific guidelines. This structure forces the LLM to move beyond superficial pattern matching to explicit, step-by-step deduction grounded in the holistic context. Finally, \tool aggregates results from multiple query rounds via majority voting to ensure detection robustness.}

\textbf{Evaluation.} 
We extensively evaluated \tool on effectiveness, ablation, generality, efficiency, and usefulness. On the \textsc{PrimeVul4J} dataset, \tool achieved a precision of \todo{\revision{0.82}}, recall of \todo{\revision{0.71}}, and F1-score of \todo{\revision{0.76}}, exceeding the best learning-based, LLM-based, and agent-based approaches by \todo{\revision{25}}\%, \todo{\revision{17}}\%, and \todo{\revision{21}}\% in F1-score, respectively. Notably, \tool attained a VP-S score of \todo{\revision{0.58}}, surpassing the best baseline by \todo{\revision{164}}\% and demonstrating robust capability to distinguish vulnerable code from patched versions.
Our ablation study validates that each component contributes to \tool's effectiveness, with holistic vulnerability context being the most critical. For generality, \tool achieved an F1-score of \todo{0.78} on the C/C++ \textsc{PrimeVul} dataset with minimal adaptation, consistently outperforming all baselines.
Regarding usefulness, \tool detected \todo{26} true vulnerabilities across \todo{9} real-world Java projects, with \todo{15} confirmed and \todo{5} CVE identifiers assigned. In industrial deployment, \tool identified \todo{40} confirmed vulnerabilities.

\textbf{Contributions.} This work makes the following main contributions.
\begin{itemize}[leftmargin=*, nosep]
    \item \textit{\textbf{Novel Approach.}} We propose \tool, a vulnerability detection approach that leverages neuro-symbolic strategies to construct an enhanced program dependency graph. This graph serves as a rigorous foundation for holistic vulnerability context extraction, which guides context-aware LLM reasoning to achieve effective vulnerability detection.
    \item \textit{\textbf{Extensive Evaluation.}} We conduct comprehensive experiments to indicate that \tool consistently outperforms learning-based, LLM-based, and agent-based approaches on Java \textsc{PrimeVul4J} dataset, and achieves strong generality on the C/C++ \textsc{PrimeVul} dataset.
    \item \textit{\textbf{Practical Impact.}} We apply \tool to \todo{9} real-world Java projects, detecting \todo{26} true vulnerabilities with \todo{15} confirmed by developers and \todo{5} CVE identifiers assigned, along with \todo{40} additional confirmed vulnerabilities in an industrial deployment.
\end{itemize}


\section{Motivation} \label{sec:motivation}
\revision{We highlight the two key limitations of current vulnerability detection approaches through two real-world false positive cases and a pilot study, motivating the need for \tool.}

\begin{figure}[!t]
    \centering
    \includegraphics[width=0.8\textwidth]{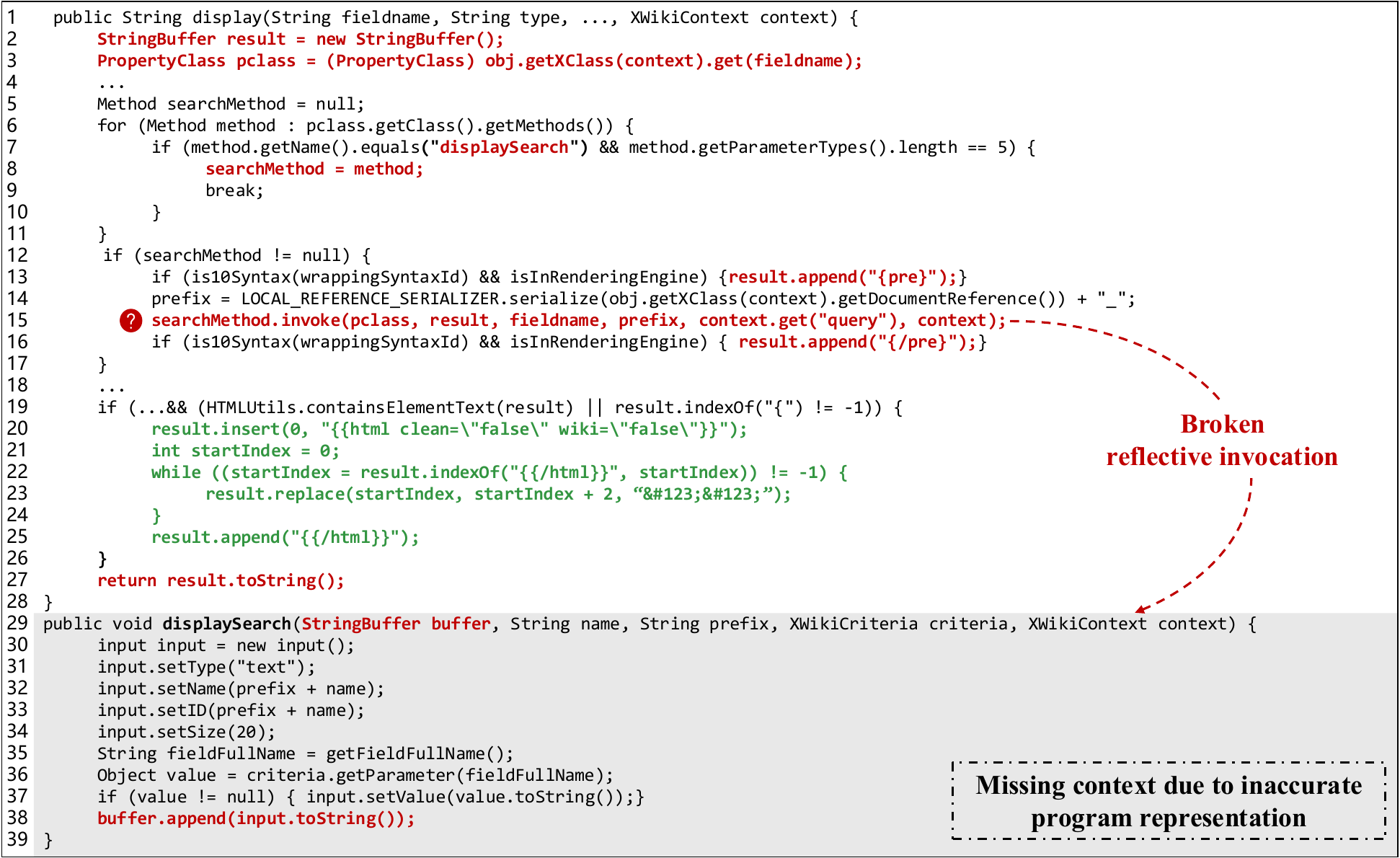}
    \vspace{-5pt}
    \caption{Patched Version of CVE-2023-29523}
    \label{fig:sample_case1}
\end{figure}

\revision{\textbf{Example 1: Incomplete Vulnerability Context Due to Inaccurate Program Representations.}} 
Fig.~\ref{fig:sample_case1} illustrates a representative false positive arising from the inaccurate program representation. This is the patched version of CVE-2023-29523~\cite{CVE-2023-29523nvd}. Here, the \texttt{display} method invokes \texttt{displaySearch} via reflection (Line 15), which populates the \texttt{result} buffer with user input (Lines 30–38). Before being returned for rendering (Line 27), this buffer is sanitized (Lines 20–25) to prevent macro injection.

\revision{However, constructing an accurate call graph for such dynamic language features is a known challenge~\cite{landman2017challenges}. Scalable static analysis tools typically fail to resolve the target of the reflective call \texttt{searchMethod.invoke}, resulting in an under-approximation of the call graph that breaks the call edge between \texttt{display} and \texttt{displaySearch}.}
As a result, approaches dependent on these inaccurate representations~\cite{lekssays2025llmxcpg,lu2024grace} restrict the LLM to the scope of caller, obscuring the callee \texttt{displaySearch}. Although the LLM can infer that the reflective call targets \texttt{displaySearch}, it cannot access its implementation. Faced with this ambiguity, the model hallucinates that the reflective invocation introduces complex malicious payloads that could bypass local sanitization, thereby erroneously flagging the method as vulnerable.
\revision{This misclassification illustrates the first facet of \textbf{Limitation 1}, i.e., inaccurate program representations lead to incomplete vulnerability context, causing the LLM to reason over missing information.}


\begin{figure}[!t]
    \centering
    \includegraphics[width=0.8\textwidth]{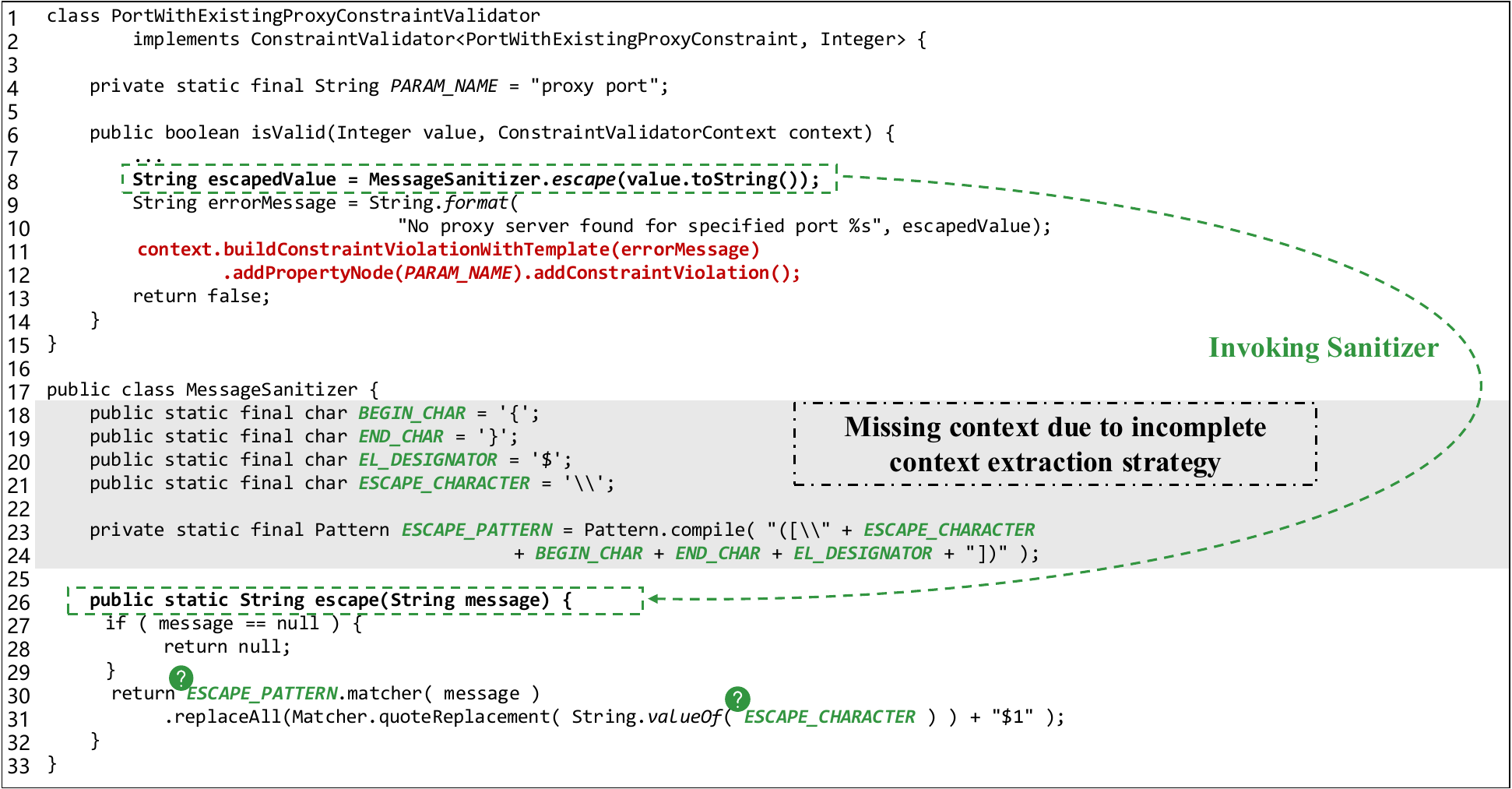}
    \vspace{-5pt}
    \caption{Patched Version of CVE-2020-26282}
    \label{fig:sample_case2}
\end{figure}

\revision{\textbf{Example 2: Incomplete Vulnerability Context Due to Insufficient Context Extraction.}} 
Fig.~\ref{fig:sample_case2} illustrates a representative false positive arising from insufficient context extraction. This is the patched version of CVE-2020-26282~\cite{CVE-2020-26282nvd}. 
At first glance, the sensitive invocation in the \texttt{isValid} method (Line~11) appears vulnerable because attacker-controlled input \texttt{value} is passed to the sensitive \texttt{buildConstraintViolationWithTemplate} invocation, which performs message interpolation that could evaluate injected Java Expression Language (EL) fragments, potentially leading to server-side template injection and remote code execution.
However, the vulnerability is mitigated by the sanitization call \texttt{escape} at Line 8. Crucially, this sanitizer relies on a specific regular expression initialized in the static field \texttt{ESCAPE\_PATTERN} (Lines 18-24) to identify and neutralize potential EL fragments, thereby eliminating security risks.

However, existing approaches struggle to capture this mitigation because their analyses are typically function-centric. Most approaches~\cite{widyasari2025let, zhu2025specification} limit analysis to the body of \texttt{isValid}, omitting the sanitizer. More importantly, even approaches that support inter-procedural slicing, such as \textsc{LLMxCPG}~\cite{lekssays2025llmxcpg}, typically misclassify this sample as vulnerable.  
While these approaches can trace inter-procedural paths into the \texttt{escape} body, their slicing algorithms are restricted to explicit control and data dependencies, overlooking implicit context such as global variable definitions. 
The core sanitization in \texttt{escape} depends on the definition of \texttt{ESCAPE\_PATTERN}, which, being structurally separated from data flow, is omitted from the extracted context. 
Consequently, the LLM cannot verify the sanitization logic without the specific regex definition. In the absence of this critical information, the model adopts an over-conservative security assumption, erroneously concluding that \texttt{isValid} is vulnerable, thereby leading to a false positive.
\revision{This demonstrates the second facet of \textbf{Limitation 1}, i.e., even when the program representation is accurate, insufficient context extraction strategies omit semantically critical information, again leaving the LLM with an incomplete vulnerability context.}

\revision{\textbf{Pilot Study on Unstructured LLM Reasoning.}} \label{pilot-robustness}
\revision{We conducted a pilot study to assess how unstructured LLM reasoning impacts robustness in vulnerability detection, using the test set of \textsc{PrimeVul4J} dataset (detailed in Sec.~\ref{sec:evaluation-setup}).}
\cyh{LLMs are provided with the holistic context extracted by \tool, not isolated code snippets. We compare two prompting strategies, i.e., (1) Chain-of-Thought (CoT) prompting, which asks the LLM to ``think step by step'' and has been shown to be the optimal prompting strategy for vulnerability detection~\cite{ding2024vulnerability}; and (2) meta-prompting~\cite{zhang2023meta, suzgun2024meta}, which decomposes the detection task into a deterministic sequence of vulnerability type specific expert guidelines (e.g., dataflow tracing and control-dependency verification), thereby anchoring LLM reasoning in concrete program semantics rather than open-ended deliberation.
Both strategies receive the same holistic context, so any performance gap directly isolates the effect of reasoning structure, i.e., whether explicit, guideline-driven decomposition can prevent LLMs from relying on superficial lexical cues.}
We executed this evaluation by designing an adversarial setting where misleading semantic signals are injected through identifier renaming. Specifically, we prepended \texttt{non\_vulnerable} to identifiers in vulnerable code and \texttt{vulnerable} to those in non-vulnerable code. This semantic-preserving transformation yielded two distinct datasets, i.e., the \textit{original dataset} (\textsc{PrimeVul4J}) and the \textit{adversarial dataset} (\textsc{PrimeVul4J} w/ misleading signals).

\begin{table}[!t]
  \centering
  \scriptsize
  \caption{Robustness Results for LLMs under CoT Prompting and Meta-Prompting}
  \label{tab:llm_robustness}
  \vspace{-10pt}
  \resizebox{\textwidth}{!}{%
  \begin{tabular}{c c c c c c c c c}
    \toprule
    \multirow{4}{*}{\textbf{LLM}} & \multicolumn{4}{c}{\textbf{CoT prompting}} & \multicolumn{4}{c}{\textbf{Meta-prompting}} \\
    \cmidrule(lr){2-5} \cmidrule(lr){6-9}
    & \multicolumn{2}{c}{original dataset} & \multicolumn{2}{c}{adversarial dataset ($\Delta$)}
    & \multicolumn{2}{c}{original dataset} & \multicolumn{2}{c}{adversarial dataset ($\Delta$)} \\
    \cmidrule(lr){2-3} \cmidrule(lr){4-5} \cmidrule(lr){6-7} \cmidrule(lr){8-9}
    
      & F1 & VP-S & F1 ($\Delta$) & VP-S ($\Delta$)
      & F1 & VP-S & F1 ($\Delta$) & VP-S ($\Delta$) \\
    \midrule

    \textsc{GPT-5-mini}
      & \basecell{0.70} & \basecell{0.25}
      & \adcell{40}{0.62~{(-0.08)}} & \adcell{60}{0.14~{(-0.11)}}
      & \basecell{0.76} & \basecell{0.55}
      & \adcell{12}{0.74~{(-0.02)}} & \adcell{15}{0.52~{(-0.03)}} \\

    \textsc{Claude-Haiku-4.5}
      & \basecell{0.72} & \basecell{0.16}
      & \adcell{75}{0.60~{(-0.12)}} & \adcell{98}{-0.09~{(-0.25)}}
      & \basecell{0.73} & \basecell{0.53}
      & \adcell{12}{0.71~{(-0.02)}} & \adcell{20}{0.49~{(-0.04)}} \\

    \textsc{Deepseek-V3.2}
      & \basecell{0.65} & \basecell{0.06}
      & \adcell{40}{0.55~{(-0.10)}} & \adcell{90}{-0.07~{(-0.13)}}
      & \basecell{0.76} & \basecell{0.58}
      & \adcell{10}{0.74~{(-0.02)}} & \adcell{10}{0.57~{(-0.01)}} \\

    \textsc{Gemini-2.5-flash}
      & \basecell{0.57} & \basecell{0.04}
      & \adcell{80}{0.43~{(-0.14)}} & \adcell{98}{-0.38~{(-0.42)}}
      & \basecell{0.71} & \basecell{0.45}
      & \adcell{6}{0.71~{(0.00)}} & \adcell{35}{0.39~{(-0.06)}} \\

    \textsc{Grok-Code-Fast}
      & \basecell{0.72} & \basecell{0.21}
      & \adcell{98}{0.44~{(-0.28)}} & \adcell{98}{-0.16~{(-0.37)}}
      & \basecell{0.71} & \basecell{0.46}
      & \adcell{20}{0.67~{(-0.04)}} & \adcell{45}{0.39~{(-0.07)}} \\

    \bottomrule
  \end{tabular}%
  }
\end{table}

We assessed the robustness of five representative vanilla LLMs, i.e., \textsc{GPT-5-mini}~\cite{gpt}, \textsc{Claude-Haiku-4.5}~\cite{claude}, \textsc{Deepseek-V3.2}~\cite{deepseek}, \textsc{Gemini-2.5-flash}~\cite{gemini}, and \textsc{Grok-Code-Fast}~\cite{grok}, on vulnerability detection.
To rigorously assess LLM robustness, we report the F1-score and VP-S score. F1-score reflects the overall detection performance, while VP-S score measures pairwise robustness by subtracting the proportion of reversed predictions from the correct discriminations between vulnerable and the corresponding non-vulnerable patched code. \cyh{The metrics are detailed in Sec.~\ref{sec:evaluation-setup}.}

The left columns of Table~\ref{tab:llm_robustness} clearly demonstrate the fragility of standard CoT prompting. All evaluated LLMs show marked performance drops on the adversarial dataset, with F1-score declining by \todo{11\%} to \todo{39\%} and VP-S score dropping by \todo{44\%} to \todo{1,050\%}, even turning negative in extreme cases. This sharp degradation indicates that LLMs predominantly depend on superficial lexical cues rather than reasoning over the control or data flow semantics of the program.

In contrast, under the meta-prompting strategy in \tool, which incorporates vulnerability type specific expert guidelines, the models demonstrate remarkable resilience against adversarial perturbations. As shown in the right-hand columns of Table~\ref{tab:llm_robustness}, the performance of LLMs remains stable across original and adversarial datasets, with F1-score declines ranging from \todo{0\%} to \todo{6\%} and VP-S reductions kept within \todo{16\%}. This confirms that structured expert guidance effectively shields model reasoning against misleading lexical cues. \revision{The stark disparity between the fragility of standard CoT and the stability of meta-prompting empirically demonstrates \textbf{Limitation 2},} \cyh{underscoring that structured, vulnerability-aware reasoning guidance is indispensable for robust LLM-based detection.}

\section{Approach}

\begin{figure*}[!t]
    \centering
    \includegraphics[width=0.75\textwidth]{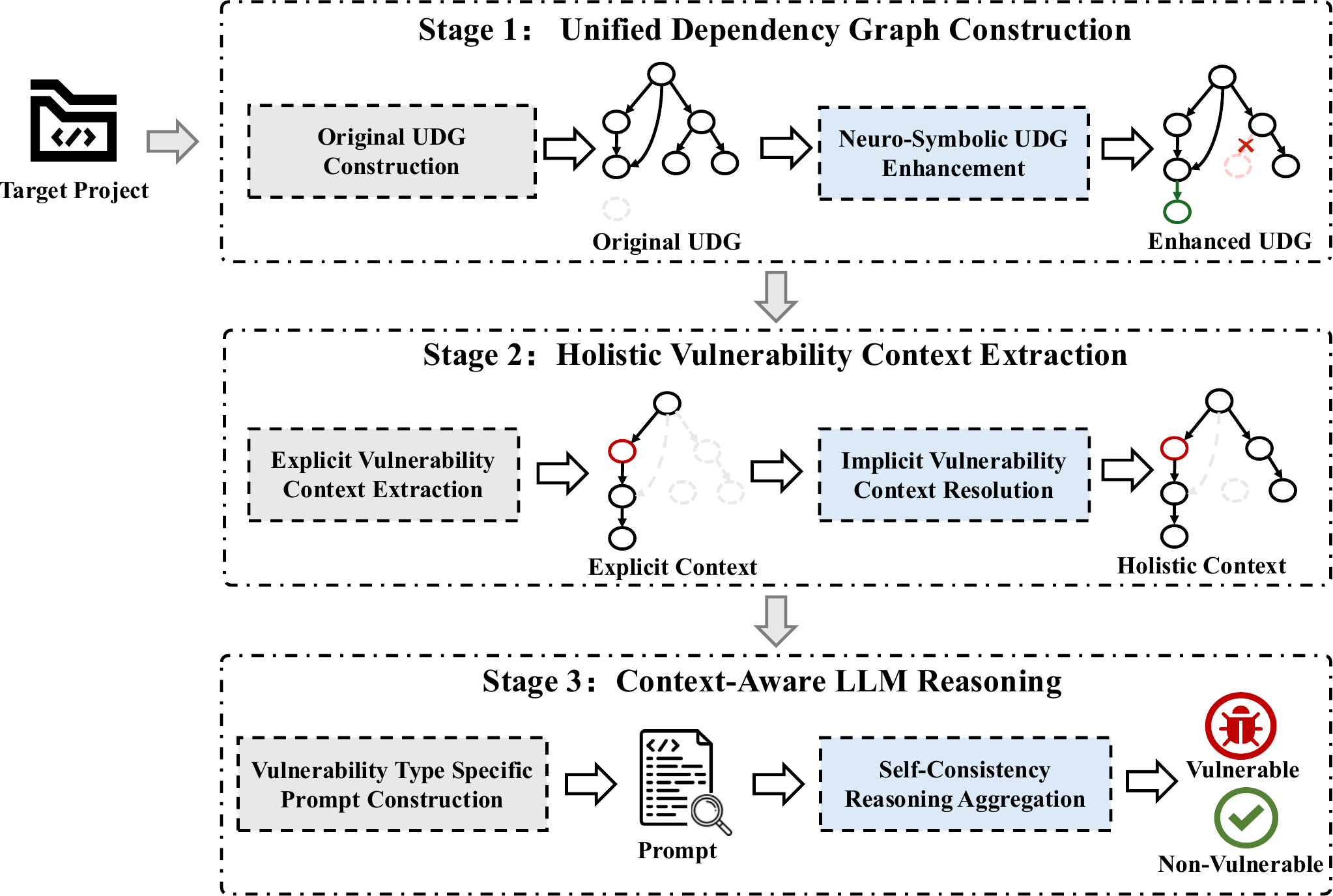}
    \vspace{-5pt}
    \caption{Overview of \tool}
    \label{fig:overview}
\end{figure*}

\subsection{Overview} \label{sec:overview}
To overcome the limitations outlined in Sec.~\ref{sec:intro}, we present \tool, a novel vulnerability detection approach. As shown in Fig.~\ref{fig:overview}, \tool comprises three main components:

\begin{enumerate}[leftmargin=1.5em, nosep]
    \item \emph{Unified Dependency Graph Construction.} Given a target repository $repo$, \tool first constructs an original unified dependency graph (UDG) $\mathcal{G}_o$ by unifying the call graph, control flow graph, and data dependency graph constructed by source code level static analysis tool. To address the limitations of static analysis, \tool further enhances this graph using heuristic rules and LLM, resulting in an enhanced, more accurate UDG $\mathcal{G}_e$ that serves as a reliable foundation for vulnerability context extraction.
    \item \emph{Holistic Vulnerability Context Extraction.} After constructing the enhanced UDG, \tool identifies sensitive invocations, which encompass predefined sensitive API invocations and, optionally, user-specified dangerous functions, and extracts holistic vulnerability contexts for each.
    Specifically, \tool first extracts the explicit context $\mathcal{C}_e$ by performing backward and forward slicing along explicit edges in $\mathcal{G}_e$, collecting all statements with direct dependencies on the sensitive invocation.
    However, such directional slicing fails to cross structural boundaries, missing essential contexts such as global variable definitions. 
    To address this, based on the explicit context $\mathcal{C}_e$, \tool employs an implicit context resolution strategy to extract implicit context $\mathcal{C}_i$ which contains statements structurally disconnected from the primary slice yet semantically necessary for analyzing the sensitive invocation. The final holistic context $\mathcal{C}$ is defined as $\mathcal{C}_e \cup \mathcal{C}_i$, ensuring completeness for each sensitive invocation.
    \item \emph{Context-Aware LLM Reasoning.} For each sensitive invocation, \tool identifies its vulnerability type, either via a predefined API mapping (e.g., \texttt{Runtime.exec()} $\rightarrow$ \texttt{CWE-78: command injection}) or a user-provided designation for custom dangerous functions, and retrieves the corresponding expert guideline. It then combines this guideline with the holistic context $\mathcal{C}$ to construct a meta-prompt~\cite{zhang2023meta, suzgun2024meta}, which is submitted to the LLM for analysis. 
    To ensure the stability of the detection results and mitigate stochastic variations in LLM outputs, \tool aggregates predictions from multiple independent queries via majority voting.
\end{enumerate}

This work focuses on Java, a language notable for its dynamic features~\cite{jdk11_doc} and widespread industrial adoption~\cite{github_octoverse}. 
This work is conducted in collaboration with a leading technology company. Our knowledge base, which maps sensitive APIs to vulnerability types and type-specific detection guidelines, was initialized by an internal tool and refined by two senior security experts, each with over five years of experience. 
\cyh{The automated initialization of the knowledge base completes within \todo{137} seconds, and extending it to a new CWE type requires only a lightweight expert review of the generated template, making the overall construction and maintenance cost practical for incremental adoption.}
Currently, this curated knowledge base encompasses \todo{102} CWE types. With the approval of our collaborator, we will release it to support reproducibility~\cite{opensource}.

\subsection{Neuro-Symbolic Construction of Unified Dependency Graph (UDG)}\label{sec:udg-construction}
\revision{As discussed in Sec.~\ref{sec:intro}, the program representation produced by static analysis serves as the foundation for vulnerability context extraction. However, scalable static analysis tools rely on conservative approximations that inevitably introduce inaccuracies into this foundation. Importantly, these inaccuracies are not specific to any single tool but stem from fundamental challenges shared across static analysis in general, such as resolving runtime-dispatched calls, modeling non-local control flow, and precisely tracking inter-procedural data dependencies. \tool addresses these challenges through a neuro-symbolic enhancement strategy that combines deterministic structural rules with LLM-based semantic inference, evolving the original representation into a more accurate foundation for vulnerability context extraction.}
\subsubsection{Original UDG Construction}
\revision{To construct the unified dependency graph (UDG), \tool first leverages Joern~\cite{joern}, a widely adopted open-source static analysis tool in vulnerability detection research~\cite{lekssays2025llmxcpg,lu2024grace,zhou2019devign}, to construct a code property graph (CPG) for the input repository $repo$.} The CPG consists of semantic representations, including the control flow graph (CFG), data dependency graph (DDG), and call graph (CG). The DDG and CFG capture intra-procedural data and control flows between the statements within functions, while the CG encodes inter-procedural links by connecting call statements with their target function entry statements. Complementing these semantic representations, the CPG also includes the abstract syntax tree (AST), which preserves the hierarchical syntactic structure of each statement.

To construct the original UDG, \tool selectively unifies only the semantic representations (i.e., DDG, CFG, and CG) extracted from the CPG, thereby focusing on data dependencies, control flows, and method invocation relationships, while omitting the syntactic details provided by the AST to avoid excessive graph complexity.
Specifically, \tool abstracts the code into statement-level nodes and preserves the underlying data dependency, control flow, and call edges, thereby unifying intra- and inter-procedural relationships. 
The result is $\mathcal{G}_o = (V_o, E_o)$, where $V_o$ is the set of statements in $repo$, and $E_o$ contains directed edges $(u, v, \tau)$, with $u, v \in V_o$ and $\tau \in \{\texttt{control flow},\allowbreak\texttt{data dependency},\allowbreak\texttt{call}\}$ indicating the type of relationship.

\subsubsection{Neuro-Symbolic UDG Enhancement} \label{sec:udg-enhancement}
The original UDG $\mathcal{G}_o$ suffers from both under-approximation (e.g., missing global definitions) and over-approximation (e.g., spurious dependencies) due to inherent limitations of static analysis. To address these issues, \tool implements a neuro-symbolic enhancement strategy that improves UDG~accuracy.
Specifically, \tool broadens the semantic scope by integrating global context into the graph structure. Then it restores control-flow continuity by enhancing call edges via LLM-based inference and refining control-flow edges through deterministic rules.
Finally, it purifies data dependencies via rigorous, summary-based analysis to eliminate spurious flows. This holistic transformation evolves $\mathcal{G}_o$ into an enhanced UDG $\mathcal{G}_e$, thereby establishing a more accurate and reliable foundation for subsequent vulnerability context extraction.

\textbf{Global Node Enhancement.} 
By focusing predominantly on function bodies, Joern inadvertently omits essential global context, including global variable definitions, \texttt{import} declarations and \texttt{package} declarations, as well as top-level class structures~\cite{jdk11_doc}. 
Recognizing this deficiency, \tool utilizes Tree-sitter~\cite{tree-sitter} to parse the input repository $repo$ and directly adds explicit nodes representing these global statements in the UDG, thereby establishing a more complete semantic scope.
After adding global nodes, \tool constructs data dependency edges only within global variable assignments among the global nodes. For each global variable assignment, it identifies variables on the right-hand side (RHS) and links the definitions of these variables to the corresponding assignment targets with data dependency edges.
To avoid unnecessary graph complexity, \tool defers creating data dependency edges from global nodes to statements inside function bodies until implicit context extraction (Sec.~\ref{sec:implicit}). Control flow and call edges for global nodes are omitted, as global statements are purely definitional and do not drive execution flow. This targeted edge construction ensures semantic completeness for vulnerability reasoning while keeping the graph compact and manageable.

\textbf{Call Edge Enhancement.} 
Static analysis often fails to accurately resolve call edges for dynamic language features, resulting in inaccurate call graphs~\cite{huangprofmal}. 
While incorporating dynamic runtime information can improve accuracy, this strategy is impractical for compiled languages like Java and C/C++ due to the significant overhead of setting up runtime environments and handling complex external dependencies~\cite{yu2025cxxcrafter}.
Recent studies~\cite{yuan2025cg} demonstrate that LLMs can significantly improve call graph construction. Therefore, \tool first heuristically identifies inaccurate call edges in $\mathcal{G}_o$ and then applies LLM to enhance them. Since polymorphism and reflection are the primary sources of call graph inaccuracy in Java according to JDK 11 Documentation~\cite{jdk11_doc}, \tool specifically targets call edges associated with these features for enhancement.

\begin{figure*}[!t]
\footnotesize
    \begin{tcolorbox}[title=Prompt of polymorphic call edge enhancement, size=fbox, left=1pt, right=1pt, top=0pt, 
      bottom=0pt,
      before upper={
      \footnotesize
      \setlength{\parskip}{0pt}
      \setlength{\parindent}{0pt}
      \linespread{0.90}\selectfont
      }]
      \cyh{\textbf{\#\#\# Task} \\ As an expert in object-oriented static analysis, analyze the provided dataflow context and identify which candidate methods are feasible targets for this specific polymorphic call statement. \\[0.75pt]
        \textbf{\#\#\# Inputs} \\
        \textbf{Dataflow Context: } The data flow context of the corresponding polymorphic call statement. \\
        \textbf{Polymorphic Call Statement: } The code of the polymorphic call statement. \\
        \textbf{Candidate Callee Method Signatures: } The list of candidate callee method signatures. \\
        \textbf{Class Inheritance Hierarchy: } The class inheritance hierarchy. 
      }
    \end{tcolorbox}
    \vspace{-10pt}
    \caption{Prompt of Polymorphic Call Edge Enhancement}
    \label{fig:polymorphism_prompt}
\end{figure*}

\begin{figure*}[!t]
\footnotesize
\begin{minipage}[t]{0.48\textwidth}
    \centering
    \vspace{0pt}
    \begin{tcolorbox}[title=Prompt of reflection call edge enhancement (Step 1), size=fbox, left=1pt, right=1pt, top=0pt, 
      bottom=0pt,
      before upper={
      \footnotesize
      \setlength{\parskip}{0pt}
      \setlength{\parindent}{0pt}
      \linespread{0.90}\selectfont
      }]
      \cyh{\textbf{\#\#\# Task} \\ As a software security expert, analyze the provided dataflow context and determine which class is accessed through this reflection API call.\\[0.75pt]
        \textbf{\#\#\# Inputs} \\
        \textbf{Dataflow Context: } The data flow context of corresponding reflection API call statement. \\
        \textbf{Reflection API Call Statement: } The code of reflection API call statement. \\
        \textbf{Available Classes: } The list of all class names in analyzed codebase.}
    \end{tcolorbox}
    \vspace{-5pt}
    \subcaption{Step 1: Identifying Target Class}
    \label{fig:reflection_step1}
\end{minipage}
\hfill
\begin{minipage}[t]{0.48\textwidth}
    \centering
    \vspace{0pt}
    \begin{tcolorbox}[title=Prompt of reflection call edge enhancement (Step 2), size=fbox, left=1pt, right=1pt, top=0pt, 
      bottom=0pt,
      before upper={
      \footnotesize
      \setlength{\parskip}{0pt}
      \setlength{\parindent}{0pt}
      \linespread{0.90}\selectfont
      }]
      \cyh{\textbf{\#\#\# Task} \\ Given the target class identified in Step 1, determine which specific method is being invoked through reflection. \\[0.75pt]
        \textbf{\#\#\# Inputs} \\
        \textbf{Dataflow Context: } The data flow context of the corresponding reflection API call statement. \\
        \textbf{Reflection API Call Statement: } The code of reflection API call statement. \\
        \textbf{Target Class Methods: } The list of all methods in identified class.}
    \end{tcolorbox}
    \vspace{-5pt}
    \subcaption{Step 2: Identifying Target Method}
    \label{fig:reflection_step2}
\end{minipage}
\vspace{-10pt}
\caption{Prompts for Reflection Call Edge Enhancement}
\label{fig:reflection_prompts}
\end{figure*}

\emph{Call Edge Enhancement for Polymorphism.} 
Joern conservatively connects each polymorphic call site to all methods with matching signatures in the inheritance hierarchy, resulting in numerous infeasible call edges that introduce irrelevant code into vulnerability detection.
To address this, \tool identifies polymorphic call statements (e.g., \texttt{a.f()}) with ambiguous dispatch targets (e.g., \texttt{a}). Since dispatch depends on the runtime type of the receiver object which can be inferenced by the dataflow context of the call statement, \tool conducts a backward data dependency analysis to precisely trace the definition and updates of the receiver object. 
In parallel, \tool leverages the pre-computed type definitions from Joern to traverse the class inheritance hierarchy and collect all candidate method signatures matching the callee signature of the call statement, thereby constructing a complete semantic context.
For each polymorphic call statement, \tool constructs an LLM prompt with: (1) the dataflow context of the receiver, (2) the call statement, (3) candidate method signatures, and (4) class inheritance hierarchy. LLM rigorously analyzes these elements to precisely determine feasible dispatch targets. Prompt templates are as shown in Fig.~\ref{fig:polymorphism_prompt}.
Guided by the LLM's decisions, \tool prunes all infeasible call edges in $\mathcal{G}_o$. This selection process retains only the edges leading to valid dispatch targets that match the runtime type of the receiver object as inferred from its dataflow context.

\emph{Call Edge Enhancement for Reflection.} Reflection defers the resolution of invoked methods or constructors until runtime, making accurate call graph construction difficult for Joern. To address this, \tool also utilizes LLMs to enhance the call edges for reflection. Since reflective calls usually invoke APIs in \texttt{java.lang.reflect}~\cite{jdk11_doc}, \tool first identifies all statements invoking these APIs. 
For each such call statement, \tool performs backward data dependency analysis to recover the definitions and modification histories of all arguments passed to the reflection call, and then constructs an LLM prompt comprising: (1) the dataflow context of these arguments, (2) the reflection API call statement itself, and (3) the repository-wide list of class names.
LLM determines which class is targeted by the reflection call. \tool then extracts the method signatures of this class, and constructs a second LLM prompt to identify the specific method invoked. Guided by LLM responses, \tool removes the original call edges to the reflection API and adds new edges to the resolved target methods. Both prompts are as shown in Fig.~\ref{fig:reflection_prompts}.

\revision{To mitigate potential LLM hallucinations in both polymorphism and reflection enhancement, \tool requires the LLM to produce a confidence score (ranging from 0 to 10) alongside each decision. Only when the score exceeds a configurable threshold $\theta_c$ (set to $8$ by default) does \tool accept the recommendation. Otherwise, the original call edges remain unchanged. We evaluate the sensitivity of $\theta_c$ in Sec.~\ref{sec:parameter-sensitivity}.}

\textbf{Control Flow Edge Enhancement.} 
Accurate construction of CFG for intra-procedural non-local jumps is a major challenge in scalable static analysis. Joern captures only structured intra-procedural control flow, leaving non-local jumps, such as labeled \texttt{break} and \texttt{continue} statements, often unresolved. This results in discontinuities in control-flow paths within $\mathcal{G}_o$, leading to critical context loss during vulnerability detection.
To address this, \tool implements a deterministic reconstruction mechanism for labeled jumps. Using Tree-sitter~\cite{tree-sitter}, \tool identifies the target of each labeled \texttt{break} or \texttt{continue} statement. Specifically, labeled \texttt{continue} can only target iteration constructs (i.e., \texttt{while}, \texttt{do-while}, \texttt{for}, \texttt{for-each}), while labeled \texttt{break} may target any labeled statement, including iteration constructs, \texttt{switch} statements, and arbitrary labeled blocks. 

Upon resolving the target, \tool deterministically reconstructs the missing CFG edges. For labeled \texttt{continue}, the jump is connected to next-iteration point of the loop (i.e., the update statement of \texttt{for} and \texttt{for-each} loops, or the condition check of \texttt{while} and \texttt{do-while} loops). For labeled \texttt{break}, the jump targets the statement immediately after the relevant labeled construct. This strategy effectively restores non-local control flow paths lacking in standard analysis.
\revision{We scope this enhancement to labeled jumps because their targets are syntactically explicit and can therefore be resolved deterministically. Other intra-procedural control-flow constructs, such as exception handling and try-with-resources, are already modeled correctly by the underlying static analysis framework. Asynchronous callbacks, which introduce implicit cross-function control flow, remain an open challenge and are discussed in Sec.~\ref{sec:discussion}.}

\textbf{Data Dependency Edge Enhancement.} \label{data-enhancement}
Scalable static analysis tools like Joern use conservative heuristics for inter-procedural data dependencies, treating every function argument as potentially affecting the return value at each call site. This over-approximation introduces many spurious data dependency edges in $\mathcal{G}_o$. To eliminate these false dependencies, \tool performs a bottom-up, summary-based data flow analysis over the enhanced call and control flow edges, enforcing accurate flow constraints and significantly improving dependency accuracy.

Concretely, \tool first partitions $\mathcal{G}_o$ into a set of functions $F = \{f_1, f_2, \dots, f_n\}$, each associated with its internal statements and intra-procedural edges. For any function $f_i \in F$, let $\mathcal{P}_i$ denote the set of its formal parameters and $r_i$ its return value. \tool then summarizes the data dependencies within $f_i$ by computing a mapping $\Phi_i: \mathcal{P}_i \to \{0, 1\}$, where:
\begin{equation}
\Phi_i(p) =
\begin{cases}
1 & \text{if } r_i\ \text{data-depends on } p \\
0 & \text{otherwise}
\end{cases}
\end{equation}
Thus, $\Phi_i(p) = 1$ when the parameter $p$ influences $r_i$. All data dependency edges in $\mathcal{G}_o$ are then pruned based on these summaries, yielding a more accurate graph.

\textit{Function Ordering.} 
To ensure correctness, \tool must analyze each function only after all its callees have been summarized, so that callee summaries are available when processing each caller. This demands a processing order corresponding to a valid topological sort of the call graph. However, recursion and mutual recursion introduce cycles, making a strict partial order infeasible.

To this end, \tool first abstracts a Function Call Graph (FCG) from $\mathcal{G}_o$, where each node represents a function in $F$ and each edge signifies an inter-procedural call relationship. To handle recursion, \tool condenses this FCG into a Directed Acyclic Graph (DAG) by grouping mutually recursive functions into single components.
Specifically, \tool applies Tarjan's algorithm~\cite{Tarjan} to partition the FCG into strongly connected components (SCCs), denote as $\mathcal{S} = \{S_1, S_2, \dots, S_k\}$, where each $S_i$ is a maximal set of mutually reachable functions (i.e., a single function or a group of mutually recursive functions). This condensation replaces cycles in the original graph with SCC nodes, yielding a DAG structure that ensures the feasibility of topological processing for mutually dependent functions.
Finally, by performing a reverse topological sort on this condensation graph, \tool derives the analysis sequence $\mathcal{Q} = \{S_{q_1}, \dots, S_{q_m}\}$. This sequence guarantees that before analyzing any component $S_{q_i}$, all its callees have been fully processed, ensuring sound summary propagation even in the presence of recursion.

\textit{Component-Wise Summary Generation.}
Given the analysis sequence $\mathcal{Q}$, \tool computes summaries for each component $S_{q_i}$ based on its topological properties. For atomic components (i.e., single non-recursive functions), the summary is derived in a single intra-procedural pass. In contrast, recursive components require fixed-point iteration, where the summaries of all functions within the component are iteratively updated until convergence is achieved.

Regardless of whether single-pass or fixed-point iteration is used, the core analysis for each function is identical. \tool formulates the intra-procedural derivation for an arbitrary function $f \in F$ as a taint-style reachability analysis, where the \textit{taint status} of a variable signifies whether it is dependent on the formal parameters.
Specifically, \tool initiates the propagation by marking all formal parameters as the initial taint sources. Rather than relying on the original, inaccurate data dependency edges, it propagates taint solely along the enhanced control flow edges in $\mathcal{G}_o$. To ensure flow sensitivity, \tool employs a worklist algorithm~\cite{dataflow} that iteratively refines taint status along these paths. For each statement, specialized transfer functions are applied:
\begin{itemize}[leftmargin=1em, nosep]
    \item \emph{Assignments:} Taint status is propagated from right-hand side (RHS) expressions to the left-hand side (LHS) variable. To handle reference types, \tool uses Tree-sitter~\cite{tree-sitter} to identify reference assignments and build alias sets grouping variables that point to the same memory location. When a variable's taint status changes, \tool propagates the update to every member of its alias set, keeping taint consistent across aliased references.
    \item \emph{Method Invocations:} If the callee $f_{callee}$ is defined within the repository (i.e., $f_{callee}\in F$), its summary $\Phi_{callee}$ is applied. For external callees (i.e., $f_{callee}\notin F$), taint is conservatively assumed to flow from all arguments to the return value.
    \item \emph{Return Statements:} The taint status of the return variable is recorded and aggregated to define the function summary $\Phi_f$, precisely reflecting statement-level data dependencies.
\end{itemize}

\textit{Data Dependency Edge Pruning.} After obtaining function summaries, \tool prunes data dependency edges in $\mathcal{G}_o$ using the computed summary information. For each call site $s_{call}$ invoking $f_{callee} \in F$, \tool matches each argument $x$ to its corresponding formal parameter $p \in \mathcal{P}_{callee}$ and consults $\Phi_{callee}(p)$ to determine whether a true dependency should exist. Only if $\Phi_{callee}(p) = 1$ does \tool retain the data dependency edge from the definition of argument $x$ to the call site $s_{call}$, otherwise, the edge is eliminated. 
This strategy ensures that $\mathcal{G}_o$ contains only semantically valid data dependencies at each call site, accurately reflecting actual data flow of the program.

With all enhancements to global nodes, call edges, control flow edges, and data dependency edges applied, \tool constructs the enhanced UDG $\mathcal{G}_e = (V_e, E_e)$, where $V_e$ comprises both original and new global nodes, and $E_e$ contains all refined edges, establishing a more accurate foundation for subsequent vulnerability context extraction.
\vspace{-2pt}
\begin{exmp}\label{exmp:udg-enhancement}
\cyh{For the example in Fig.~\ref{fig:sample_case1}, \tool first identifies the \texttt{Method.invoke} call statement (Line~15), then performs backward data dependency analysis to recover the definitions and modification histories of all arguments, extracting the dataflow context (Lines~5-11) and the list of all class names in the codebase. The LLM then determines that the target class is \texttt{PropertyClass}, and with the method list of this target class, further determines that the target method is \texttt{displaySearch}, enhancing the call edge from \texttt{display} to \texttt{displaySearch}.}
\end{exmp}
\vspace{-2pt}

\subsection{Holistic Vulnerability Context Extraction} \label{sec:context-extraction}
\tool extracts the holistic vulnerability context from the enhanced UDG $\mathcal{G}_e$.
\subsubsection{Explicit Vulnerability Context Extraction}
Given a target repository $repo$, \tool traverses the enhanced UDG $\mathcal{G}_e$ to collect all sensitive API invocations, denoted as $\mathcal{I}$. Our complete list of sensitive APIs and their corresponding vulnerability types is publicly available at our website~\cite{opensource}. 
Beyond the built-in knowledge base, \tool allows users to specify additional dangerous functions. All invocations of these user-defined functions are automatically identified in $\mathcal{G}_e$ and incorporated into $\mathcal{I}$.
For each $i \in \mathcal{I}$ with the corresponding sensitive call statement $s_i$, \tool conducts both \emph{data dependency} and \emph{control flow} slicing on $\mathcal{G}_e$, using $s_i$ as the slicing criterion.

\textbf{Data Dependency Slicing.} For each sensitive call statement $s_i$, \tool conducts both forward and backward data dependency slicing on $\mathcal{G}_e$ with $s_i$ as the slicing criterion. This process traverses data dependency edges to collect all statements that $s_i$ depends on as well as statements dependent on $s_i$, thus capturing how variables are defined and propagated in relation to $s_i$. The resulting data dependency context is denoted as $\mathcal{C}_{data}$.

\textbf{Control Flow Slicing.} \tool performs control flow slicing on $\mathcal{G}_e$ with $s_i$ as the slicing criterion. It traverses control flow and call edges to collect statements that influence or are influenced by $s_i$, capturing both intra- and inter-procedural control flow contexts. The resulting control flow context is denoted as $\mathcal{C}_{control}$. 
\revision{To prevent excessive context and token overflow, we introduce a configurable hop limit $h$ that caps inter-procedural call edge traversal. Following prior work~\cite{li2024effectiveness}, we set $h=3$ by default.}
The final explicit vulnerability context is $\mathcal{C}_e = \mathcal{C}_{data} \cup \mathcal{C}_{control}$.
\vspace{-6pt}
\subsubsection{Implicit Vulnerability Context Resolution}\label{sec:implicit}
While explicit slicing effectively isolates the relevant semantic context, relying solely on it is insufficient to form a self-contained semantic unit for accurate vulnerability detection. 
Strictly directional slicing is inherently limited, often omitting essential semantic information beyond slice boundaries, such as intra-procedural logic within callees, upstream variable definitions excluded by slicing direction, and global declarations needed for type inference that are unconnected due to missing edges. 
To address the incomplete context issue, \tool employs a unified implicit context resolution strategy that recovers missing semantic fragments by extracting usage, definition, and declaration contexts.

\textbf{Usage Context Resolution.} 
Explicit slicing based on unidirectional traversal often yields shallow context, missing how variables are actually used or sanitized within callee functions. For example, suppose a sensitive API call statement $s_b$ (e.g., \texttt{Files.createFile(path)}) uses a variable such as \texttt{path}, which is the return value of preceding call statement $s_a$. If the callee function $f_c$ validates or sanitizes \texttt{path}, conventional backward slicing from $s_b$ will only reach $s_a$ and stop at the function boundary, omitting the internal logic implemented at $f_c$. 
As a result, crucial sanitization context is excluded, and the analysis may falsely report $s_b$ as unprotected, increasing false positives.

To mitigate this, \tool first extracts all call statements from $\mathcal{C}_e$ and get the corresponding callee entry statements. For each callee entry statement, \tool performs forward data dependency slicing on $\mathcal{G}_e$ with the callee entry statement as the slicing criterion. The resulting data dependency context is denoted as $\mathcal{C}_{use}$.

\textbf{Definition Context Resolution.} 
The initially extracted explicit context $\mathcal{C}_e$ frequently lacks complete variable definitions, referencing variables whose initializations lie outside the captured boundaries. This deficiency arises from two primary sources. First, the unidirectional initial forward slicing inherently omits upstream definitions. Second, global data dependency edges are strategically deferred to prevent graph densification (as detailed in Sec.~\ref{sec:udg-enhancement}), which structurally disconnects global definitions from their usages. To address this, \tool implements an adaptive backward recovery strategy guided by structural reachability of variable definitions.

Specifically, \tool first identifies unresolved variables as $V_{unresolved} = V_{use} \setminus V_{def}$, where $V_{use}$ and $V_{def}$ represent the variables used and defined within $\mathcal{C}_e$, respectively. For each $v \in V_{unresolved}$ at its usage statement $s_{use}$, \tool examines the incoming data dependency edges for $v$ in $\mathcal{G}_e$ to determine if a definition is connected.
If no incoming edge exists, $v$ is likely a global variable whose linkage is missing. Therefore, \tool locates $v$'s global definition $s_{def}$ and performs backward data dependency slicing from $s_{def}$ to recover its initialization history. If incoming edges are present, indicating that the definition is structurally reachable, \tool directly performs backward data dependency slicing from $s_{use}$ to recover the definition within the local scope.
All retrieved statements are finally aggregated as the definition context $\mathcal{C}_{def}$.

\textbf{Declaration Context Resolution.} 
Beyond recovering variable usages and definitions, it is equally vital to capture the global structural context required for accurate type resolution. However, the initially extracted $\mathcal{C}_e$ often misses structural global declarations, i.e., \texttt{import}, \texttt{package}, and top-level class statements, because they are disconnected from code within function bodies in the $\mathcal{G}_e$ and unreachable via standard slicing. Yet, these declarations are critical for accurate type resolution by LLMs~\cite{zhu2025specification}. Without them, type names remain ambiguous (e.g., distinguishing \texttt{java.util.List} from \texttt{java.awt.List} for the \texttt{List} usage). To address this, \tool locates the source files~related to $\mathcal{C}_e$ in $repo$ and extracts these structural declarations, forming $\mathcal{C}_{decl}$, which provides a complete and unambiguous type resolution context. The final implicit context is $\mathcal{C}_i = \mathcal{C}_{use} \cup \mathcal{C}_{def} \cup \mathcal{C}_{decl}$. 

\vspace{-2pt}
\begin{exmp}\label{exmp:implicit-context-resolution}
\cyh{For the example in Fig.~\ref{fig:sample_case2}, \tool identifies the sensitive API call \texttt{build\-Constraint\-Violation\-With\-Template} (Line~11), which may introduce an expression language injection vulnerability. The explicit context $\mathcal{C}_e$ comprises the backward slice (Lines~6, 8--10), the sensitive invocation itself (Lines~11--12), and the forward slice (Line~13), collectively covering Line~6 and Lines~8--13. Since $\mathcal{C}_e$ omits the internal sanitization logic of \texttt{Message\-Sanitizer.escape}, where \texttt{value} is transformed into \texttt{escapedValue}, \tool performs \emph{usage context resolution} to obtain $\mathcal{C}_{use}$ (Lines~26--31). The global variables \texttt{ESCAPE\_PATTERN}, \texttt{ESCAPE\_CHARACTER}, and \texttt{PARAM\_NAME} referenced in $\mathcal{C}_e$ and $\mathcal{C}_{use}$ are absent from their respective contexts, as they are modeled as global nodes in the enhanced UDG (Sec.~\ref{sec:udg-enhancement}). \tool therefore applies \emph{definition context resolution} to recover $\mathcal{C}_{def}$ (Lines~18--24 and Line~4). Finally, because structural global declarations are unreachable via standard slicing, \tool applies \emph{declaration context resolution} to obtain $\mathcal{C}_{decl}$ (Lines~1, 2, 17). The implicit context is thus $\mathcal{C}_i = \mathcal{C}_{use} \cup \mathcal{C}_{def} \cup \mathcal{C}_{decl} = \{1,\; 2,\; 4,\; 17\text{--}24,\; 26\text{--}31\}$, and the holistic context is $\mathcal{C} = \mathcal{C}_e \cup \mathcal{C}_i = \{1,\; 2,\; 4,\; 6,\; 8\text{--}13,\; 17\text{--}24,\; 26\text{--}31\}$.}
\end{exmp}
\vspace{-2pt}

\subsection{Context-Aware LLM Reasoning}~\label{sec:context-aware-llm-reasoning}

\tool detects vulnerabilities by constructing a meta-prompt that integrates the holistic~context~$\mathcal{C} = \mathcal{C}_e \cup \mathcal{C}_i$ with a vulnerability-specific guideline~\cite{zhang2023meta, suzgun2024meta}. This prompt is submitted to the LLM, and the verdict is obtained by aggregating multiple independent queries via majority voting.

\subsubsection{Vulnerability Type Specific Prompt Construction}
Fig.~\ref{fig:detection_prompt} presents the meta-prompting template~$\mathcal{P}$ for vulnerability detection. 
\cyh{Unlike vanilla Chain-of-Thought (CoT) prompting in vulnerability detection~\cite{nong2024chain, steenhoek2024comprehensive, tamberg2025harnessing, ullah2024llms, zhang2024prompt, zhou2024large}, which simply asks the LLM to reason step by step but lacks transparent and grounded reasoning (as demonstrated in Sec.~\ref{pilot-robustness}), meta-prompting introduces a structured, high-level framework. 
It enables \tool to transform abstract detection tasks into a deterministic sequence of analytical primitives, tailored to the unique semantic requirements of different vulnerability types. By explicitly decomposing the reasoning process into dataflow tracing and control-dependency verification, \tool ensures that the LLM's conclusion is anchored in concrete program semantics rather than statistical hallucinations.}
\cyh{As detailed in Sec.~\ref{sec:overview}, we provide guidelines for \todo{102} CWE types, publicly available online~\cite{opensource}. Each meta-prompt is structured in two parts, i.e., a problem description and solution criteria. This clear separation enforces a disciplined, context-aware reasoning process for the LLM.}

\begin{figure*}[t]
    \footnotesize
        \begin{tcolorbox}[title=Meta prompt template for vulnerability detection, size=fbox, left=1pt, right=1pt, top=0pt, 
          bottom=0pt,
          before upper={
          \footnotesize
          \setlength{\parskip}{0pt}
          \setlength{\parindent}{0pt}
          \linespread{0.90}\selectfont
          }]
            \textbf{\#\#\# Problem} \\ You are an expert security auditor for Java. Given a specific code context \textbf{\%$\mathcal{C}$\%} encompassing a target sensitive invocation of \textbf{\%$\mathit{API}$\%}, analyze \textbf{\%$\mathcal{C}$\%} to determine whether a genuine \textbf{\%$\mathit{cwe}$\%} vulnerability exists.\\[0.75pt]
            \textbf{\#\#\# Solution Instructions} \\
            Follow the \textbf{guideline} below step by step. After completing each step, critically review your reasoning for overlooked issues (e.g., implicit sanitization, broken dataflow, incomplete context) and revise your analysis as necessary. Continue this review until you reach a well-justified conclusion. Summarize your final answer in the following JSON format: \{"explanation": \texttt{<step-by-step reasoning>}, "is\_vulnerable": \texttt{true or false}\}.\\
            \cyh{\textbf{\#\#\# Vulnerability Type Specific Guideline}: \\
            \textbf{Step 1: Contextual Flow Understanding} -- Starting from the sensitive invocation of \textbf{\%$\mathit{API}$\%}, precisely understand and extract all relevant data and control flow paths within \textbf{\%$\mathcal{C}$\%}.\\[0.75pt]
            \textbf{Step 2: Trigger Condition Verification} -- Systematically evaluate whether the extracted paths fulfill the exact vulnerability conditions specified by \textbf{\%$vuln\_patterns_{cwe}$\%}.\\[0.75pt]
            \textbf{Step 3: Defense Assessment} -- Critically examine defense mechanisms present in $\mathcal{C}$ using \textbf{\%$defense\_knowledge_{cwe}$\%}, clearly distinguishing robust mitigations from insufficient or bypassable ones.\\[0.75pt]
            \textbf{Step 4: Evidence-Driven Verdict Synthesis} -- Synthesize prior artifacts. Weight verified exploitable indicators against active counter-evidence before deriving a final verdict.}
        \end{tcolorbox}
        \vspace{-10pt}
        \caption{Meta Prompt Template for Vulnerability Detection}
        \label{fig:detection_prompt}
    \end{figure*}
\begin{itemize}[leftmargin=1em, nosep]
    \item \textbf{Problem} clearly defines the vulnerability detection task by providing the LLM with three essential inputs, i.e., the holistic context \%$\mathcal{C}$\% of the sensitive API invocation from Sec.~\ref{sec:context-extraction}, the sensitive API \%$\mathit{API}$\% and its corresponding vulnerability type \%$\mathit{cwe}$\%. Crucially, these two elements are treated as an inseparable pair, whether they are derived from the pre-defined knowledge base or provided by the user through a unified specification that maps customized dangerous functions to their respective vulnerability types. This ensures that all necessary information is available for accurate, context-aware analysis.
    \cyh{\item \textbf{Solution} orchestrates the reasoning process via a specialized four-step, CWE-centric trajectory rather than relying on the LLM's implicit reasoning. We explicitly parameterize this guideline with two categories of domain-specific knowledge, which are initialized from an industrial auditing tool and refined through rigorous expert validation (as detailed in Sec.~\ref{sec:overview}).}
    \cyh{\begin{enumerate}[label=(\arabic*), leftmargin=2em, nosep]
        \item \textbf{Vulnerability Semantics} ($\%vuln\_patterns_{cwe}\%$) formalizes abstract vulnerability characteristics into verifiable semantic predicates for the analysis in Step 2. Specifically, it transforms high-level security violations into concrete data-flow or control-flow constraints. For instance, the semantic pattern for SQL Injection (CWE-89) is operationalized as a taint-style reachability problem, i.e., verifying whether untrusted data from an external source flows into a SQL execution without passing through proper parameterization or sanitization routines.
        \item \textbf{Remediation Heuristics} ($\%defense\_knowledge_{cwe}\%$) catalogs both robust mitigations and commonly flawed ad-hoc defenses frequently encountered in industrial audits, serving as the baseline for the \emph{Defense Assessment} in Step~3. For instance, the remediation heuristic for CWE-89 (SQL Injection) instructs the LLM to recognize \texttt{PreparedStatement} with bind variables as a robust defense, because the database driver treats every parameter as a literal value rather than executable SQL, structurally preventing injection regardless of input content. Conversely, it flags manual string-level sanitization (e.g., calling \texttt{replaceAll} to strip quote characters) as an unreliable ad-hoc defense, since such character-level filtering is inherently incomplete and easily bypassed through encoding tricks or context-dependent escape sequences.
    \end{enumerate}}
    \cyh{Grounded in this domain specific knowledge, the guideline forces the LLM to execute a rigorous sequence of \emph{Contextual Flow Understanding}, \emph{Trigger Condition Verification}, and \emph{Defense Assessment}. Ultimately, \emph{Evidence-Driven Verdict Synthesis} mandates a hypothesis-driven conclusion, i.e., the LLM must explicitly weigh exploitable indicators (e.g., valid paths, triggering logic) against counter-evidence (e.g., adequate sanitization) before deriving a final verdict. 
    By requiring each step to produce explicit intermediate evidence, this structure prevents the LLM from jumping to a conclusion without reconciling conflicting signals in the context.}
\end{itemize}

Crucially, recognizing that a single API may be susceptible to multiple vulnerability types, \tool adopts a fine-grained detection strategy. It treats each valid $(API, cwe)$ pair as a distinct detection unit, iterating through all potential vulnerability types for a given API to generate a dedicated prompt for each specific vulnerability type.

\subsubsection{Self-Consistency Reasoning Aggregation}~\label{sec:self-consistency-reasoning-aggregation}
Given a vulnerability type specific prompt $\mathcal{P}$, \tool queries the LLM $\mathcal{N}$ times independently to account for randomness. It then aggregates results using majority voting, ensuring robust and consistent vulnerability predictions~\cite{wang2022self}.

\vspace{-2pt}
\begin{exmp}\label{exmp:self-consistency-reasoning-aggregation}
\cyh{Following Example~\ref{exmp:implicit-context-resolution}, \tool constructs the meta-prompt for expression language injection (CWE-74) detection using the holistic context $\mathcal{C}$, queries the LLM $\mathcal{N}$ times independently, and aggregates the results via majority voting, yielding a final verdict of \texttt{false} and eliminating the false positive.}
\end{exmp}
\vspace{-2pt}


\section{Evaluation}\label{sec:evaluation}
We have implemented \tool with \todo{7K} lines of Python code. We used \textsc{DeepSeek V3.2}~\cite{deepseek} as the inference model in \tool. We design \todo{seven} research questions.~All~experiments were conducted on a Linux server with 256 GB RAM and an Nvidia A100 GPU (80~GB~memory).
\begin{itemize}[leftmargin=*, nosep]
    \item \textbf{RQ1 Effectiveness Evaluation.} How effective is \tool in detecting vulnerabilities?
    \item \textbf{RQ2 Ablation Study.} How is the contribution of each component in \tool?
    \item \textbf{RQ3 Parameter Sensitivity.} How do the configurable parameters affect \tool?
    \item \textbf{RQ4 Generality Evaluation.} How is the generality of \tool across different language?
    \item \textbf{RQ5 Efficiency Evaluation.} How is the efficiency of \tool?
    \item \textbf{RQ6 Usefulness Evaluation.} How is the practical usefulness of \tool?
    \item \textbf{RQ7 UDG Enhancement Evaluation.} How effective is \tool in UDG enhancement?
\end{itemize}

\subsection{Evaluation Setup}\label{sec:evaluation-setup}

\textbf{Dataset.} \cyh{Vulnerability detection suffers from substantial label noise in existing benchmarks. While \textsc{PrimeVul}~\cite{ding2024vulnerability} offers high-quality labels, it only covers C/C++, whereas our work targets Java. To address this, we curate \textsc{PrimeVul4J}, a Java vulnerability dataset closely following \textsc{PrimeVul}'s rigorous methodology~\cite{ding2024vulnerability}. The detailed steps are as follows.}
\cyh{\begin{enumerate}[leftmargin=*, nosep]
    \item \textbf{Data Collection and Labeling.} We aggregated the Java subsets of \textsc{CrossVul}~\cite{nikitopoulos2021crossvul}, \textsc{ReposVul}~\cite{wang2024reposvul}, and \textsc{CVEFixes}~\cite{bhandari2021cvefixes}, and extended the corpus with their respective collection scripts~\cite{crawlscripts, crawlscripts_reposvul, crawlscripts_CVEfixes} to include vulnerabilities disclosed up to \todo{January 31, 2026}. To mitigate the label noise inherent in automated collection, we then applied the labeling methodology proposed by \textsc{PrimeVul} \cite{ding2024vulnerability} to denoise the corpus, yielding \todo{816} vulnerable and \todo{842} non-vulnerable functions, each annotated with its CWE categories.
    \item \textbf{Normalization and Rigorous Deduplication.} To ensure rigorous evaluation and prevent data leakage via code clones, we first normalize all modified functions by thoroughly removing formatting characters (e.g., spaces, tabs, newlines, carriage returns) and compute their MD5 hashes for both pre- and post-commit versions. Functions with identical pre- and post-commit hashes are considered unchanged and discarded. Subsequently, the remaining samples undergo a global deduplication process based on their normalized hashes. Crucially, when hash collisions occur among multiple samples, we strictly retain the oldest chronological version and discard the newer ones to further eliminate the risk of temporal data leakage. This rigorous process ultimately yields \todo{1620} unique samples, comprising \todo{790} vulnerable and \todo{830} non-vulnerable functions.
    \item \textbf{Chronological Data Splitting.} To formulate a realistic evaluation setup and avert "time travel" leakage~\cite{ding2024vulnerability} for the train methods, the deduplicated dataset is partitioned strictly by commit timestamps. The oldest 80\% of samples form the training set, the next 10\% are the validation set, and the latest 10\% are the test set, keeping functions from the same commit together. This results in \todo{627} vulnerable and \todo{663} non-vulnerable training samples, \todo{79} vulnerable and \todo{83} non-vulnerable validation samples, and \todo{84} vulnerable and \todo{84} non-vulnerable test samples. Each sample has the label and its corresponding CWE categories.
    \item \textbf{Paired Subset Construction.} To rigorously assess the capability of a model in distinguishing subtle semantic differences between vulnerable and non-vulnerable code, we pair each vulnerable function with its post-fix version from the same commit, yielding \todo{675} unique pairs split into \todo{528} for training, \todo{75} for validation, and \todo{72} for testing. The pair count is smaller than the individual sample count because some commits delete vulnerable functions entirely or introduce entirely new non-vulnerable ones, leaving no matching counterpart.
\end{enumerate}
}
\cyh{A central challenge in our evaluation is the granularity mismatch, while \textsc{PrimeVul4J} provides ground truth at the function level, \tool operates at the repository level to leverage holistic context.
To resolve this mismatch, we input the entire repository containing the target function into \tool and extract its repository-wide holistic context for detection. Specifically, we dynamically determine the slicing criteria for context extraction based on the function's characteristics. If the target function invokes a sensitive API from our knowledge base~\cite{opensource}, this invocation naturally serves as the slicing criterion. Otherwise, we treat the target function itself as a dangerous sink; we identify all its call sites across the repository and use these statements as the slicing criteria. 
In both scenarios, we augment the prompt with vulnerability-specific guidelines, ensuring that \tool's reasoning aligns strictly with the detection criteria of the target vulnerability type.}

\begin{figure*}[!t]
    \centering
    \includegraphics[width=0.85\textwidth]{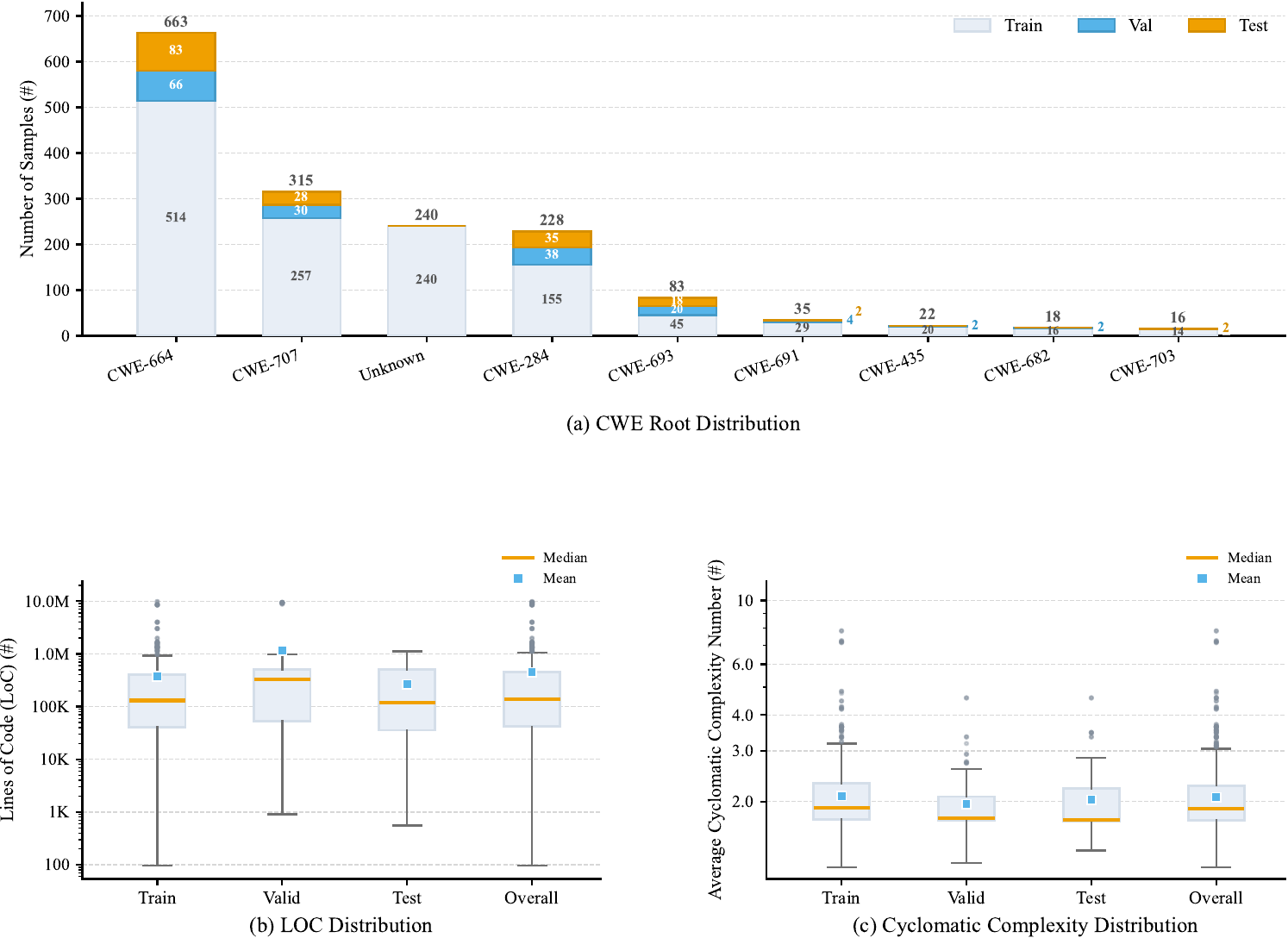}
    \vspace{-5pt}
    \caption{\revision{Vulnerability Type Distribution, and Project Scale and Complexity Distribution in \textsc{PrimeVul4J}}}
    \label{fig:dataset_combined_characteristics}
\end{figure*}

\revision{\textbf{Characteristics of the Dataset.} To demonstrate the diversity of \textsc{PrimeVul4J}, we analyze its code-level and vulnerability type distributions, as shown in Fig.~\ref{fig:dataset_combined_characteristics}.}

\revision{\emph{Vulnerability Type Distribution.} \textsc{PrimeVul4J} covers \todo{102} distinct CWE types in total. Following the classifications in the official CWE hierarchical views~\cite{cwe-view-of-research-concepts, cwe-view-of-software-development}, we categorize these CWEs into \todo{9} broad groups based on their common ancestors. As shown in Fig.~\ref{fig:dataset_combined_characteristics}(a), these CWE types are distributed across all 9 top-level categories, with CWE-664 and CWE-707 being the most prevalent. Notably, the MITRE CWE Top 25~\cite{cwe-top25} spans 7 of our 9 categories, indicating that \textsc{PrimeVul4J} aligns well with the most critical vulnerability classes in practice.}

\revision{\emph{Code Scale and Complexity.} Fig.~\ref{fig:dataset_combined_characteristics}(b) presents the project-level LoC distribution across the dataset. The projects exhibit substantial scale variation, with a median of approximately \todo{224K} LoC and a mean of approximately \todo{719K} LoC, while the largest projects exceed \todo{7M} LoC, demonstrating that our dataset encompasses projects ranging from moderate-sized libraries to large-scale industrial systems. Furthermore, Fig.~\ref{fig:dataset_combined_characteristics}(c) presents the average cyclomatic complexity distribution per project. The median is approximately \todo{1.9} with a mean of approximately \todo{2.1}, while outliers reach up to \todo{7}, indicating that the dataset covers projects with varying levels of structural complexity. Notably, the distributions are consistent across all three splits, confirming that our chronological partitioning does not introduce distributional~shift.}

\cyh{\textbf{Baseline Selection.} We compare \tool with representative approaches spanning three categories. 
(1) For learning-based approaches, we select \textsc{DeepDFA}~\cite{steenhoek2024dataflow}. Although primarily evaluated on C/C++, it fundamentally relies on Joern, a language-agnostic static analysis framework. Therefore, we seamlessly adapt it to Java and retrain it on our \textsc{PrimeVul4J} for a fair comparison. 
(2) For LLM-based detection, we select baselines covering three dominant paradigms, i.e., fine-tuning, Retrieval-Augmented Generation (RAG), and direct prompting. 
For fine-tuning, we select \textsc{LLMxCPG}~\cite{lekssays2025llmxcpg}. Similarly, leveraging its dependence on Joern, we extend its open-source implementation for Java and fine-tune it using their publicly available scripts~\cite{llmxcpg}. 
For RAG, \textsc{VulInstruct}~\cite{zhu2025specification} and \textsc{VulRAG}~\cite{du2024vul} are selected. 
For direct prompting, we evaluate the standard Chain-of-Thought (\textsc{CoT})~\cite{ding2024vulnerability} to represent unaugmented LLM reasoning. Besides, we include \textsc{IRIS}~\cite{li2024iris}, which augments LLM prompts with interprocedural taint paths extracted via traditional static analysis, making it the most closely related baseline to our approach. 
Finally, we exclude \textsc{LLMDFA} from our evaluation, as its source and sink identification mechanism is strictly limited to only three vulnerability types, and scaling it to a broader range of CWEs would require prohibitive manual effort.
\revision{(3) For agent-based approaches, we select \textsc{VulTrial}~\cite{widyasari2025let} as a representative multi-agent baseline. We additionally reproduce the ReAct-style single-agent from \textsc{JitVul}~\cite{yildiz2025benchmarking} to cover the single-agent paradigm. \textsc{CPRVul}~\cite{li2026beyond} is excluded because its source code is not publicly available. \textsc{RepoAudit}~\cite{guo2025repoaudit} is excluded because its implementation natively supports only three vulnerability types, and extending it to the diverse CWEs in our dataset would require substantial manual engineering.}}
All baselines are evaluated with their default hyperparameters. For baselines that do not require training, we use \textsc{DeepSeek V3.2} as the inference model for fair comparison with \tool. \cyh{However, for the fine-tuning baseline (\textsc{LLMxCPG}), fine-tuning a massive model like \textsc{DeepSeek V3.2} incurs prohibitive computational overhead. Therefore, we utilize \textsc{Qwen2.5-32B}, a widely available~open-weights~model demonstrating comparable reasoning performance, as the base model for~fine-tuning~and~subsequent inference.}

\textbf{Evaluation Metrics.} \revision{We adopt the standard metrics widely used in vulnerability detection~\cite{ding2024vulnerability,steenhoek2024dataflow,zhu2025specification}. Specifically, we record one True Positive (TP) when the approach correctly predicts a vulnerable sample as vulnerable, one False Positive (FP) when the approach incorrectly predicts a non-vulnerable sample as vulnerable, and one False Negative (FN) when the approach fails to identify an actually vulnerable sample. Based on these counts, we compute:
\begin{equation}
  \text{Pre.} = \frac{\text{TP}}{\text{TP}+\text{FP}}, \quad
  \text{Rec.} = \frac{\text{TP}}{\text{TP}+\text{FN}}, \quad
  \text{F1} = \frac{2 \cdot \text{Pre.} \cdot \text{Rec.}}{\text{Pre.} + \text{Rec.}}
\end{equation}
In practice, precision is particularly important for vulnerability detection, because excessive false positives impose a prohibitive triage burden and cause alert fatigue~\cite{du2026reducing}, while recall ensures that genuine vulnerabilities are not missed.}

\revision{While the above metrics assess per-sample accuracy, they evaluate each sample independently, and do not measure whether an approach can correctly identify both the presence and absence of vulnerabilities in a textually similar context. To complement this, since each vulnerability fix commit naturally yields a pre-patch (vulnerable) and post-patch (non-vulnerable) version of the same function, we pair them to form matched evaluation units, and adopt pairwise metrics~\cite{wen2025boosting,ding2024vulnerability} that evaluate predictions on each such pair as a single entity. Let $N$ denote the total number of pairs, and let $(\hat{y}_v, \hat{y}_p) \in \{0, 1\}^2$ denote the predicted labels for the vulnerable and patched samples within a pair. The following metrics are defined and measured.
\begin{itemize}[leftmargin=*, nosep]
    \item \textbf{Pairwise Correct (P-C)}: the proportion of pairs where an approach correctly predicts both elements, i.e., $\text{P-C} = |\{(\hat{y}_v, \hat{y}_p) = (1, 0)\}| / N$.
    \item \textbf{Pairwise Reversed (P-R)}: the proportion of pairs where an approach inversely predicts the labels, i.e., $\text{P-R} = |\{(\hat{y}_v, \hat{y}_p) = (0, 1)\}| / N$.
\end{itemize}
The remaining two outcomes, $(\hat{y}_v, \hat{y}_p) \in \{(1,1), (0,0)\}$, represent partial predictions where exactly one element is correct. Since they are neither fully correct nor fully reversed, they are excluded from both P-C and P-R. Building on these two components, we define:
\begin{itemize}[leftmargin=*, nosep]
    \item \textbf{VP-S}: $\text{VP-S} = \text{P-C} - \text{P-R}$, which credits correct pairwise discrimination and penalizes reversed predictions. A VP-S of 1.0 means every pair is correctly distinguished, while a negative value means reversed predictions outnumber correct ones.
\end{itemize}
For instance, given 10 vulnerable--patched pairs, suppose a detector correctly distinguishes both versions in 6 pairs, reverses the labels in 1 pair, and makes partial predictions (i.e., $(1,1)$ or $(0,0)$) in the remaining 3 pairs. Then P-C\,=\,0.6, P-R\,=\,0.1, and VP-S\,=\,0.5. This example illustrates that VP-S measures pairwise discriminative capability, i.e., whether a detector can correctly distinguish between textually similar vulnerable and patched code rather than merely performing well on isolated samples.
We report both standard and pairwise metrics throughout our evaluation to provide a comprehensive assessment.}


\subsection{Effectiveness Evaluation (RQ1)}

\begin{table*}[!t]
    \centering
    \caption{Results of Our Effectiveness Evaluation and Ablation Study (\textsc{DeepDFA} and \textsc{LLMxCPG} require training, and thus their results on the whole dataset are not available (--) to avoid data leakage)}
    \label{tab:effectiveness_and_ablation_comparison}
    \vspace{-10pt}
    \renewcommand{\arraystretch}{1.1}    \setlength{\tabcolsep}{4pt}
    \resizebox{\textwidth}{!}{%
    \begin{tabular}{lcccccccccccccc}
      \toprule
      \multirow{4}{*}{\textbf{Category}} & \multirow{4}{*}{\textbf{Approach}} & \multirow{4}{*}{\textbf{Model}} & \multicolumn{6}{c}{\textbf{Test Set of \textsc{PrimeVul4J}}} & \multicolumn{6}{c}{\textbf{Whole \textsc{PrimeVul4J} Dataset}} \\
      \cmidrule(lr){4-9} \cmidrule(lr){10-15}
      & & & \multicolumn{3}{c}{\textbf{Standard Metrics}} & \multicolumn{3}{c}{\textbf{Pairwise Metrics}} & \multicolumn{3}{c}{\textbf{Standard Metrics}} & \multicolumn{3}{c}{\textbf{Pairwise Metrics}} \\
      \cmidrule(lr){4-6} \cmidrule(lr){7-9}\cmidrule(lr){10-12} \cmidrule(lr){13-15}
      & & & Pre. & Rec. & F1 & P-C & P-R & VP-S & Pre. & Rec. & F1 & P-C & P-R & VP-S \\
      \midrule
      \multirow{1}{*}{\textbf{Learning-Based}} 
      & \textsc{DeepDFA} & \textsc{GNN-based} & 0.57 & 0.66 & 0.61 & 0.25 & 0.20 & 0.05 & -- & -- & -- & -- & -- & -- \\
      \midrule
      \multirow{2}{*}{\textbf{Agent-Based}} 
      & \textsc{VulTrial} & \textsc{DeepSeek V3.2} & \todo{0.50} & \todo{0.44} & \todo{0.47} & \todo{0.16} & \todo{0.12} & \todo{0.04} & 0.51 & 0.37 & 0.43 & 0.20 & 0.15 & 0.05 \\
      & \textsc{JitVul} & \textsc{DeepSeek V3.2} & \revision{0.55} & \revision{0.74} & \revision{0.63} & \revision{0.24} & \revision{0.10} & \revision{0.14} & \revision{0.54} & \revision{0.78} & \revision{0.64} & \revision{0.25} & \revision{0.09} & \revision{0.16} \\
      \midrule
      \multirow{6}{*}{\textbf{LLM-Based}} & \textsc{CoT} & \textsc{DeepSeek V3.2} & 0.59 & 0.45 & 0.51 & 0.24 & 0.07 & 0.17 & 0.56 & 0.40 & 0.47 & 0.09 & 0.02 & 0.07 \\
      & \textsc{IRIS} & \textsc{DeepSeek V3.2} & \todo{0.22} & \todo{0.56} & \todo{0.32} & \todo{0.04} & \todo{0.01} & \todo{0.03} & 0.27 & 0.53 & 0.36 & 0.07 & 0.02 & 0.05 \\
      & \textsc{LLMxCPG} & \textsc{Qwen2.5-32B} & 0.63 & 0.04 & 0.08 & 0.30 & 0.25 & 0.05 & -- & -- & -- & -- & -- & -- \\
      & \textsc{VulRAG} & \textsc{DeepSeek V3.2} & \todo{0.56} & \todo{0.58} & \todo{0.57} & \todo{0.25} & \todo{0.03} & \todo{0.22} & 0.57 & 0.47 & 0.52 & 0.23 & 0.04 & 0.19 \\
      & \textsc{VulInstruct} & \textsc{DeepSeek V3.2} & \todo{0.53} & \textbf{\todo{0.84}} & \todo{0.65} & \todo{0.08} & \todo{0.02} & \todo{0.06} & 0.51 & \textbf{0.87} & 0.64 & 0.13 & 0.03 & 0.10 \\
      & \textbf{\tool} & \textsc{DeepSeek V3.2} & \textbf{\todo{\revision{0.82}}} & \todo{\revision{0.71}} & \textbf{\revision{0.76}} & \textbf{\todo{\revision{0.58}}} & \textbf{\todo{\revision{0.00}}} & \textbf{\todo{\revision{0.58}}} & \textbf{\revision{0.79}} & \revision{0.71} & \textbf{\revision{0.75}} & \textbf{\revision{0.55}} & \textbf{\revision{0.01}} & \textbf{\revision{0.54}} \\
      \cmidrule(lr){1-15}
      \multirow{5}{*}{\textbf{Ablations}} & w/o $\mathcal{G}_e$ & \textsc{DeepSeek V3.2} & \todo{0.70} & \todo{0.67} & \todo{0.68} & \todo{0.44} & \todo{0.02} & \todo{0.42} & 0.67 & 0.63 & 0.65 & 0.30 & 0.03 & 0.27 \\
      & w/o $\mathcal{C}_i$ & \textsc{DeepSeek V3.2} & \todo{\revision{0.76}} & \todo{\revision{0.53}} & \todo{\revision{0.62}} & \todo{\revision{0.38}} & \todo{\revision{0.00}} & \todo{\revision{0.38}} & \revision{0.74} & \revision{0.56} & \revision{0.64} & \revision{0.43} & \revision{0.02} & \revision{0.41} \\
      & w/o $\mathcal{C}$  & \textsc{DeepSeek V3.2} & \todo{0.63} & \todo{0.58} & \todo{0.60} & \todo{0.38} & \todo{0.01} & \todo{0.37} & 0.69 & 0.51 & 0.59 & 0.26 & 0.02 & 0.24 \\
      & w/ \textsc{CoT} & \textsc{DeepSeek V3.2} & \todo{\revision{0.52}} & \todo{\revision{0.86}} & \todo{\revision{0.65}} & \todo{\revision{0.11}} & \todo{\revision{0.05}} & \todo{\revision{0.06}} & \revision{0.53} & \revision{0.84} & \revision{0.65} & \revision{0.12} & \revision{0.03} & \revision{0.09} \\
      & w/ \textsc{Qwen} & \textsc{Qwen2.5-32B} & \todo{\revision{0.80}} & \todo{\revision{0.69}} & \todo{\revision{0.74}} & \todo{\revision{0.54}} & \todo{\revision{0.02}} & \todo{\revision{0.52}} & \revision{0.78} & \revision{0.71} & \revision{0.74} & \revision{0.51} & \revision{0.02} & \revision{0.49} \\
      \bottomrule
    \end{tabular}%
    }
  \end{table*}

\textbf{RQ1 Setup.} We evaluate all baselines on \textsc{PrimeVul4J} dataset. As \textsc{DeepDFA} and \textsc{LLMxCPG} require training, we train or fine-tune them on the designated training and validation sets, and report their results only on the test set to prevent data leakage. The remaining approaches are assessed on the entire \textsc{PrimeVul4J} dataset, and we report results on both the test set and the whole dataset. 
\cyh{Beyond overall effectiveness, a robust vulnerability detector should generalize well across diverse vulnerability patterns. Therefore, we further conduct a fine-grained analysis on the test set of \textsc{PrimeVul4J} to investigate model performance across different CWE types. Specifically, the test set includes \todo{168} samples encompassing \todo{41} distinct CWEs. To ensure a statistically meaningful comparison, we categorize these CWEs into \todo{6} broad groups based on their common ancestors, following the classifications in the official CWE hierarchical views~\cite{cwe-view-of-research-concepts, cwe-view-of-software-development}}.
\cyh{Lastly, to investigate the ``token tax'' (i.e., the potential performance degradation caused by the \textit{lost-in-the-middle} phenomenon~\cite{liu2024lost} as context length increases), we categorize the \textsc{PrimeVul4J} test set based on the token count within \tool's holistic context. Notably, we employ the tokenizer of DeepSeek-V3.2 to calculate these costs, allowing us to analyze the effectiveness of \tool with respect to context length.}

\textbf{Overall Results.} The upper part of Table~\ref{tab:effectiveness_and_ablation_comparison} summarizes the effectiveness evaluation. On the test set, \tool outperforms all baselines across standard and pairwise metrics, achieving a precision of \todo{\revision{0.82}}, recall of \todo{\revision{0.71}}, and F1-score of \todo{\revision{0.76}}. Compared to the leading learning-based approach \textsc{DeepDFA}, \tool improves precision by \todo{\revision{0.25}} (\todo{\revision{44}}\%), recall by \todo{\revision{0.05}} (\todo{\revision{8}}\%), and F1-score by \todo{\revision{0.15}} (\todo{\revision{25}}\%). \revision{Against the strongest agent-based baseline, \textsc{JitVul}, \tool increases precision by \todo{\revision{0.27}} (\todo{\revision{49}}\%), and F1-score by \todo{\revision{0.13}} (\todo{\revision{21}}\%). }
\revision{\cyh{For LLM-based approaches, \tool outperforms the state-of-the-art baseline, \textsc{VulInstruct}, by \todo{\revision{0.11}} (\todo{\revision{17}}\%) in F1-score. Although \textsc{VulInstruct} and \textsc{JitVul} achieve higher recall, its low precision reflects a tendency toward over-prediction. In practice, precision is often more critical than recall for vulnerability detection, as excessive false positives impose a prohibitive triage burden and lead to alert fatigue~\cite{du2026reducing}.} \tool achieves a more favorable trade-off by maintaining competitive recall while markedly reducing false positives, yielding more actionable findings with significantly lower verification cost.}
\textsc{LLMxCPG} exhibits extremely low recall because its generated CPGQL queries for Joern often fail to extract adequate context or generate correct queries. Conversely, \tool analyzes the enhanced UDG, derived from Joern (see Sec.~\ref{sec:udg-construction}), yielding a \todo{\revision{1,675}}\% higher recall.
These results demonstrate the superior effectiveness of \tool for vulnerability detection.

In terms of pairwise metrics, \tool exhibits exceptional discriminative power. 
It achieves a VP-S of \todo{\revision{0.58}} on the test set, outperforming the strongest baseline \textsc{VulRAG} by \todo{\revision{164}}\%. 
Remarkably, \tool achieves a P-R of \todo{\revision{0.00}} on the test set, underscoring its ability to perceive subtle semantic shifts rather than being misled by superficial similarities between vulnerable and non-vulnerable patched code.

On the whole \textsc{PrimeVul4J} dataset, \tool exhibits exceptional stability, maintaining an F1-score of \todo{\revision{0.75}} and a VP-S score of \todo{\revision{0.54}}. The small performance gap between the test set and the whole dataset indicates that \tool remains stable across different evaluation scales and generalizes well to diverse vulnerability patterns.

\cyh{
  \begin{table*}[!t]
  \footnotesize
 \centering
 \caption{Results of Our Effectiveness Evaluation w.r.t. CWE Categories (\#S. denotes the number of samples in a CWE category)}
 \vspace{-10pt}
 \label{tbl:cwe}
 \begin{tabular}{
 lccccccc
}
   \toprule
   \multicolumn{2}{c}{\multirow{2}{*}{Approach}} & \rotatebox[origin=c]{0}{\sc{CWE-664}} & \rotatebox[origin=c]{0}{\sc{CWE-284}} & \rotatebox[origin=c]{0}{\sc{CWE-707}} & \rotatebox[origin=c]{0}{\sc{CWE-693}} & \rotatebox[origin=c]{0}{\sc{CWE-691}} & \rotatebox[origin=c]{0}{\sc{CWE-703}} \\
   \cmidrule(lr){3-3}
   \cmidrule(lr){4-4}
   \cmidrule(lr){5-5}
   \cmidrule(lr){6-6}
   \cmidrule(lr){7-7}
   \cmidrule(lr){8-8}
   & & \multicolumn{1}{c}{\#S.=83} & \multicolumn{1}{c}{\#S.=35} & \multicolumn{1}{c}{\#S.=28} & \multicolumn{1}{c}{\#S.=18} & \multicolumn{1}{c}{\#S.=2} & \multicolumn{1}{c}{\#S.=2}\\
   \midrule
   \multirow{2}{*}{\sc{DeepDFA}} & F1 & 0.57 & 0.63 & 0.59 & 0.60 & 0.50 & 0.79 \\
   & VP-S & 0.04 & 0.07 & 0.05 & 0.03 & 0.06 & 0.09 \\
   \multirow{2}{*}{\sc{VulTrial}} & F1 & 0.44 & 0.40 & 0.46 & 0.63 & 1.00 & 0.67 \\
   & VP-S & -0.08 & -0.06 & 0.00 & 0.33 & 1.00 & 0.00 \\
   \multirow{2}{*}{\sc{CoT}} & F1 & 0.57 & 0.70 & 0.65 & 0.29 & 0.67 & 0.67 \\
   & VP-S & 0.03 & 0.31 & 0.21 & 0.00 & 0.00 & 0.00 \\
   \multirow{2}{*}{\sc{IRIS}} & F1 & 0.22 & 0.40 & 0.31 & 0.27 & 0.00 & 0.00 \\
   & VP-S & 0.05 & -0.03 & -0.01 & 0.02 & 0.00 & 0.00 \\
   \multirow{2}{*}{\sc{LLMxCPG}} & F1 & 0.03 & 0.09 & 0.07 & 0.17 & 0.00 & 0.00 \\
   & VP-S & 0.01 & 0.07 & 0.05 & 0.09 & 0.00 & 0.00 \\
   \multirow{2}{*}{\sc{VulRAG}} & F1 & 0.54 & 0.61 & 0.69 & 0.29 & 1.00 & 0.67 \\
   & VP-S & 0.21 & 0.27 & 0.36 & 0.00 & 1.00 & 0.00 \\
   \multirow{2}{*}{\sc{VulInstruct}} & F1 & 0.68 & 0.70 & 0.67 & 0.62 & 0.67 & 0.67 \\
   & VP-S & 0.11 & 0.13 & 0.00 & 0.00 & 0.00 & 0.00 \\
   \multirow{2}{*}{\sc{JitVul}} & F1 & \revision{0.62} & \revision{0.68} & \revision{0.58} & \revision{0.67} & \revision{0.00} & \revision{1.00} \\
   & VP-S & \revision{0.13} & \revision{0.24} & \revision{0.00} & \revision{0.22} & \revision{-1.00} & \revision{1.00} \\
   \multirow{2}{*}{\sc{VulWeaver}} & F1 & \textbf{\revision{0.69}} & \textbf{\revision{0.91}} & \textbf{\revision{0.78}} & \textbf{\revision{0.71}} & \textbf{\revision{1.00}} & \textbf{\revision{1.00}} \\
   & VP-S & \textbf{\revision{0.45}} & \textbf{\revision{0.80}} & \textbf{\revision{0.67}} & \textbf{\revision{0.33}} & \textbf{\revision{1.00}} & \textbf{\revision{1.00}} \\
   \bottomrule
 \end{tabular}
\end{table*}
}
\cyh{\textbf{Effectiveness w.r.t. CWE Categories.} Table~\ref{tbl:cwe} presents a fine-grained breakdown across six CWE \todo{categories}. \tool achieves the highest F1 score in all \todo{6} categories. For pairwise metrics, it strictly outperforms all baselines in three categories (i.e., CWE-664, CWE-284, and CWE-707), and ties for the best in the other three categories (i.e., CWE-693, CWE-691, and CWE-703). The ties in CWE-691 and CWE-703 should be interpreted cautiously due to their extremely small sample size (\#S.=2). These results confirm that \tool generalizes robustly across diverse vulnerability semantics.}

\begin{figure*}[!t]
  \centering

  \begin{minipage}[t]{0.24\textwidth}
    \vspace{0pt}
    \centering
    \includegraphics[width=\linewidth]{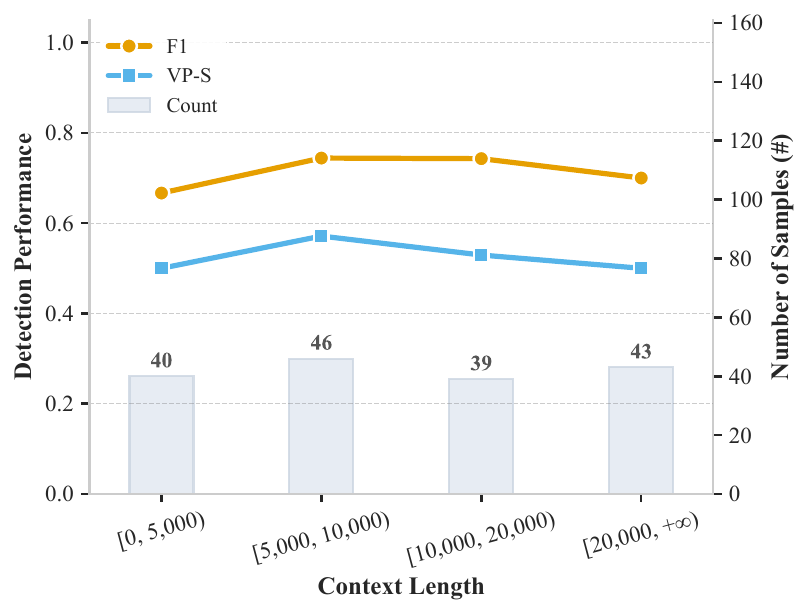}
    \par\vspace{4pt}
    \captionof{figure}{Effectiveness w.r.t. Context Length}
    \label{fig:vc-length}
  \end{minipage}
  \hfill
  \begin{minipage}[t]{0.74\textwidth}
    \vspace{0pt}
    \centering

    \begin{minipage}[t]{0.32\linewidth}
      \vspace{0pt}
      \centering
      \includegraphics[width=\linewidth]{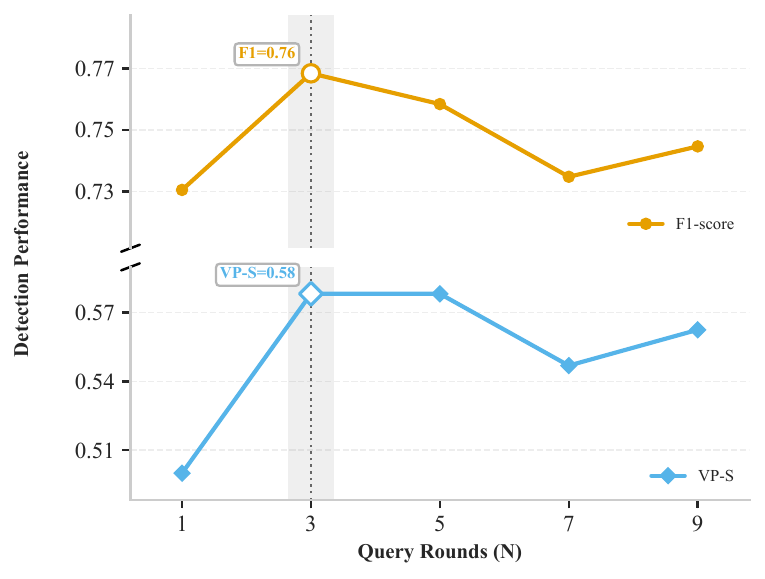}
    \end{minipage}
    \hfill
    \begin{minipage}[t]{0.32\linewidth}
      \vspace{0pt}
      \centering
      \includegraphics[width=\linewidth]{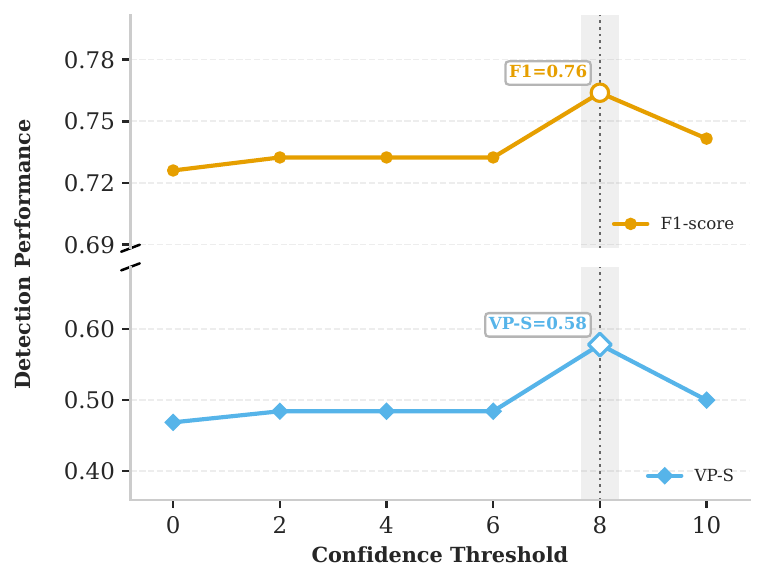}
    \end{minipage}
    \hfill
    \begin{minipage}[t]{0.32\linewidth}
      \vspace{0pt}
      \centering
      \includegraphics[width=\linewidth]{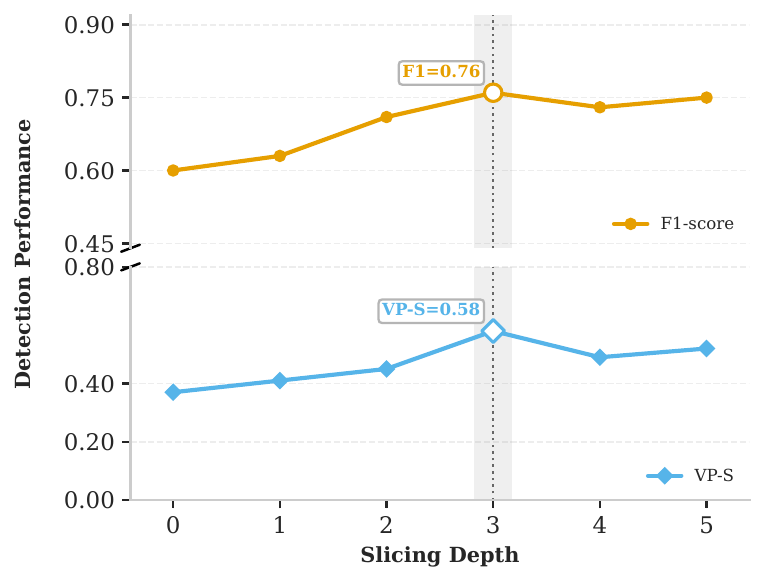}
    \end{minipage}

    \par\vspace{4pt}
    \captionof{figure}{Parameter Sensitivity Evaluation}
    \label{fig:sensitivity_rounds}
  \end{minipage}
\end{figure*}
\cyh{\textbf{Effectiveness w.r.t. Context Length.} Fig.~\ref{fig:vc-length} reports \tool's effectiveness across varying context lengths. Despite increasing token volumes, \tool maintains a high stability with only minor fluctuations, suggesting that the potential ``token tax'' from the \textit{lost-in-the-middle} phenomenon is largely mitigated by our structured context extraction. It indicates that the semantic gains from a holistic context consistently outweigh any noise-related penalties, ensuring that critical vulnerability evidence retains its decisive role in model reasoning regardless of input scale.}

\textbf{In-Depth Analysis.} 
\cyh{Overall, \tool generates \todo{\revision{148}} false positives and \todo{\revision{226}} false negatives in the whole dataset. We identify three main factors behind these misclassifications. First, source-level static analysis cannot capture opaque framework-driven behaviors in modern Java projects (e.g., Spring Boot). Declarative constructs such as dependency injection annotations introduce implicit control and data flows invisible in source code, causing \tool to produce fragmented contexts. Such fragmentation either breaks source-to-sink reachability chains, yielding false negatives, or omits global sanitization logic in framework interceptors, yielding false positives.
Second, incomplete third-party library source code forces \tool to rely on the LLM's parametric knowledge to infer external API behaviors, which may produce hallucinations, particularly for uncommon or version-dependent libraries. Consequently, the model may overestimate the risk of an API, yielding false positives, or underestimate its vulnerability propagation potential, yielding false negatives.
Finally, in a minority of cases with exceptionally extensive contexts, we observed the ``lost-in-the-middle'' phenomenon~\cite{liu2024lost} in scenarios with extensive contexts. Despite a holistic context, LLM attention sometimes fails to focus on critical but brief mitigation steps embedded within long code sequences. This can lead to both false~negatives and false positives, where the model may overlook critical mitigation logic or misinterpret benign code as~vulnerable.}

\subsection{Ablation Study (RQ2)} \label{sec:ablation}
\textbf{RQ2 Setup.} We construct \todo{five} ablated variants of \tool for analysis: (1) without UDG enhancement (w/o $\mathcal{G}_e$); (2) without implicit vulnerability context (w/o $\mathcal{C}_i$); 
(3) without holistic vulnerability context, i.e., using only the single target function as input (w/o $\mathcal{C}$); 
\cyh{(4) replacing meta-prompting with Chain-of-Thought (w/ \textsc{CoT}), using identical holistic context as the full method to isolate the effect of reasoning strategy;} and (5) substituting \textsc{DeepSeek V3.2} with \textsc{Qwen2.5-32B} (w/ \textsc{Qwen}). All ablation experiments are evaluated on the full \textsc{PrimeVul4J} dataset.
\cyh{Beyond the overall comparison, we conduct a fine-grained analysis across diverse CWE types to quantify the grounding effect of our neuro-symbolic strategy on UDG enhancement, by measuring the correlation between restored context size and the resulting gains in detection accuracy and pairwise robustness, where context size is tokenized via the DeepSeek-V3.2 tokenizer.}

\textbf{Overall Results.} The bottom part of Table~\ref{tab:effectiveness_and_ablation_comparison} presents the results of the ablation study. As observed in the whole \textsc{PrimeVul4J} dataset, the performance of \tool, measured by F1-Score and VP-S score, degrades across all five ablated versions. 
Specifically, the removal of the holistic vulnerability context (\todo{w/o $\mathcal{C}$}) causes the most significant performance drop, decreasing the F1-Score by \todo{\revision{0.16}} (\todo{\revision{21}}\%) and the VP-S score by \todo{\revision{0.30}} (\todo{\revision{56}}\%). This highlights its pivotal role in directing LLM reasoning. The absence of implicit context modeling (w/o $\mathcal{C}_i$) also leads to a notable degradation, reducing the F1-score by \todo{\revision{0.11}} (\todo{\revision{15}}\%) and the VP-S score by \todo{\revision{0.13}} (\todo{\revision{24}}\%), indicating that grasping the holistic semantics is essential for effective detection. Similarly, excluding the enhanced UDG (w/o $\mathcal{G}_e$) results in a decline in pairwise consistency, where the VP-S score drops by \todo{\revision{0.27}} (\todo{\revision{50}}\%), confirming that accurate representations are crucial for distinguishing subtle semantic nuances. 
A distinct contrast is observed with standard Chain-of-Thought (w/ \textsc{CoT}), which yields a higher recall of \todo{\revision{0.84}} relative to the \todo{\revision{0.71}} of \tool but suffers a severe precision drop by \todo{\revision{0.26}} (\todo{\revision{33}}\%). 
This result shows that without structured guidance, LLMs tend to overfit to superficial patterns, resulting in higher recall but significantly lower precision due to over-warning tendencies. \tool addresses this by introducing structured constraints, trading minimal recall for a substantial increase in precision and reliability. Moreover, substituting the backbone with \textsc{Qwen2.5-32B} (w/ \textsc{Qwen}) causes only minor performance drops.

\cyh{
\begin{table*}[!t]
  \footnotesize
  \centering
  \caption{Impact of UDG Enhancement across CWE Types}
  \vspace{-10pt}
  \label{tab:ablation-cwe}
  \begin{tabular}{
   lccccccc
  }
  \toprule
  \multicolumn{2}{c}{\multirow{2}{*}{Metric}} & \textsc{cwe-664} & \textsc{cwe-284} & \textsc{cwe-707} & \textsc{cwe-693} & \textsc{cwe-691} & \textsc{cwe-703} \\
  \cmidrule(lr){3-3} \cmidrule(lr){4-4} \cmidrule(lr){5-5} \cmidrule(lr){6-6} \cmidrule(lr){7-7} \cmidrule(lr){8-8} 
  & & \#S.=83 & \#S.=35 & \#S.=28 & \#S.=18 & \#S.=2 & \#S.=2 \\
  \midrule
  \multirow{3}{*}{Context Size} & Before & \todo{4,865} & \todo{13,254} & \todo{7,907} & \todo{1,292} & \todo{8,696} & \todo{2,323} \\
  & After & \todo{\revision{6,335}} & \todo{\revision{18,955}} & \todo{\revision{9,462}} & \todo{\revision{1,601}} & \todo{\revision{9,209}} & \todo{\revision{2,430}} \\
  & $\Delta$ (\%) & \todo{\revision{+30\%}} & \todo{\revision{+43\%}} & \todo{\revision{+20\%}} & \todo{\revision{+24\%}} & \todo{\revision{+6\%}} & \todo{\revision{+5\%}} \\
  \midrule
  \multirow{3}{*}{F1} & w/o $\mathcal{G}_e$ & \todo{0.61} & \todo{0.87} & \todo{0.53} & \todo{0.67} & \todo{1.00} & \todo{1.00} \\
  & \textbf{\tool} & \textbf{\todo{\revision{0.69}}} & \textbf{\todo{\revision{0.91}}} & \textbf{\todo{\revision{0.78}}} & \textbf{\todo{\revision{0.71}}} & \textbf{\todo{1.00}} & \textbf{\todo{1.00}} \\
 & $\Delta$ (\%) & \todo{\revision{+13\%}} & \todo{\revision{+5\%}} & \todo{\revision{+47\%}} & \todo{\revision{+6\%}} & \todo{--} & \todo{--} \\
  \midrule
  \multirow{3}{*}{VP-S} & w/o $\mathcal{G}_e$ & \todo{0.23} & \todo{0.71} & \todo{0.33} & \todo{0.33} & \todo{1.00} & \todo{1.00} \\
  & \textbf{\tool} & \textbf{\todo{\revision{0.45}}} & \textbf{\todo{\revision{0.80}}} & \textbf{\todo{\revision{0.67}}} & \textbf{\todo{\revision{0.33}}} & \textbf{\todo{1.00}} & \textbf{\todo{1.00}}  \\
  & $\Delta$ (\%) & \todo{\revision{+96\%}} & \todo{\revision{+13\%}} & \todo{\revision{+103\%}} & \todo{\revision{--}} & \todo{--} & \todo{--} \\
  \bottomrule
  \end{tabular}
\end{table*}
}
\cyh{\textbf{Impact of UDG Enhancement.} Table~\ref{tab:ablation-cwe} presents the impact of UDG enhancement on context richness and model performance across CWE types. After enhancement, context size grows by \revision{5\%} to \revision{43\%}, with the largest expansion in CWE-284 (+\revision{43\%}) and CWE-664 (+\revision{30\%}). These gains arise because access control logic in CWE-284 often involves polymorphic dispatch across authorization layers, while CWE-664 vulnerabilities routinely span cross-function allocation-to-release paths, both patterns are opaque to Joern's conservative static analysis but recovered by UDG enhancement. This structural enrichment translates into notable performance gains, with the largest improvements observed in CWE-707 (F1 +\revision{47\%}, VP-S +\revision{103\%}) and CWE-664 (VP-S +\revision{96\%}). The disproportionately large VP-S improvement in CWE-707 despite moderate context growth (+\revision{20\%}) suggests that even small amounts of recovered neutralization context are highly decisive for distinguishing vulnerable from patched code. CWE-691 and CWE-703 show negligible context growth and no change in scores, as their call graphs are already sufficiently resolved by Joern. These results empirically establish UDG enhancement as the cornerstone of \tool's effectiveness, recovering broken inter-procedural semantics to enable grounded LLM reasoning.}

\subsection{Parameter Sensitivity (RQ3)}\label{sec:parameter-sensitivity}
\revision{\textbf{RQ3 Setup.} Three parameters are configurable in \tool, i.e., the confidence threshold $\theta_c$ for accepting LLM call edge enhancement decisions (Sec.~\ref{sec:udg-enhancement}), the hop limit $h$ for inter-procedural call edge traversal (Sec.~\ref{sec:context-extraction}), and the number of query rounds $\mathcal{N}$ for self-consistency voting (Sec.~\ref{sec:self-consistency-reasoning-aggregation}). We reconfigure one parameter at a time while keeping the others fixed at their default values, and evaluate on the \textsc{PrimeVul4J} test set. $\mathcal{N}$ is varied in steps of 2 over $[1, 9]$, $\theta_c$ in steps of 1 over $[0, 10]$, and $h$ in steps of 1 over $[0, 5]$. Fig.~\ref{fig:sensitivity_rounds} reports the results.}

\textbf{Overall Results.}
\revision{As shown in Fig.~\ref{fig:sensitivity_rounds}, increasing the number of query rounds $\mathcal{N}$ from 1 to 3 improves both F1 and VP-S, with F1 reaching its peak at $\mathcal{N}=3$. Further increasing $\mathcal{N}$ yields only minor fluctuations while increasing token consumption. Therefore, $\mathcal{N}=3$ provides the best trade-off between effectiveness and efficiency, and we adopt it as the default.
For the confidence threshold $\theta_c$, both metrics improve as low-confidence LLM-based graph modifications are filtered, with the best overall performance achieved around $\theta_c=8$. When the threshold becomes too strict, performance slightly decreases because some valid enhancements may also be rejected. We therefore adopt $\theta_c=8$ as the default. 
For the slicing depth $h$, performance improves as $h$ increases from 0 to 3, where both F1 and VP-S reach their best values. This indicates that three-hop inter-procedural traversal captures sufficient vulnerability context. When $h$ is further increased, both metrics decrease, suggesting that overly deep traversal introduces irrelevant code and dilutes the context signal.
Overall, these results show that the default settings provide a stable and effective operating point, and \tool does not require extensive hyperparameter tuning in practice.}


\subsection{Generality Evaluation (RQ4)}

\textbf{RQ4 Setup.} To evaluate the cross-language generalization of \tool, we directly use the C/C++ \textsc{PrimeVul}~\cite{ding2024vulnerability} dataset. \textsc{PrimeVul} utilizes a temporal split to prevent data leakage, and its latest test set contains 435 pairs of vulnerable and non-vulnerable functions~\cite{ding2025prep}.
We directly evaluate \tool on this test set.
\revision{Since \tool is built upon Joern and Tree-sitter, both of which are multilingual, the adaptation for C/C++ is confined to a small number of changes within each component (Sec.~\ref{sec:overview}). For UDG construction (Sec.~\ref{sec:udg-construction}), we first consolidate macro definitions from included headers into a single header file, and then invoke ``\texttt{gcc~-E}'' to expand all macros before Joern parses the source, and replace the Java reflection call edge enhancement with a function pointer enhancement that reuses the same neuro-symbolic mechanism (Sec.~\ref{sec:udg-enhancement}), since both involve indirect dispatch resolved at runtime. The Java \texttt{java.lang.reflect} enhancement is omitted as it is inapplicable to C/C++. Holistic context extraction (Sec.~\ref{sec:context-extraction}) operates entirely on the UDG through graph slicing and implicit context resolution, requiring no changes. For context-aware LLM reasoning (Sec.~\ref{sec:context-aware-llm-reasoning}), the CWE guideline prescribes detection reasoning steps (e.g., taint propagation tracing, and~defense adequacy assessment) rather than concrete API names, and thus it is reused without modification. Only the~sensitive~API mappings are replaced with C/C++ counterparts (e.g., \texttt{memcpy}, \texttt{memset} for memory safety CWEs), covering~\todo{30}~CWE types. We exclude \textsc{IRIS} for comparison since it is not applicable to C/C++.}

\begin{table*}[!t]
  \centering
  \fontsize{7.0pt}{6.0pt}\selectfont
  \caption{Results of Our Generality Evaluation on the C/C++ \textsc{PrimeVul} Dataset}
  \renewcommand{\arraystretch}{1.2}
  \label{tab:generality_comparison}
  \vspace{-10pt}
  \begin{tabular*}{\textwidth}{@{\extracolsep{\fill}}lcccccccc}
    \toprule
    \multirow{2}{*}{\textbf{Category}} & \multirow{2}{*}{\textbf{Approach}} & \multirow{2}{*}{\textbf{Model}} & \multicolumn{3}{c}{\textbf{Standard Metrics}} & \multicolumn{3}{c}{\textbf{Pairwise Metrics}} \\
    \cmidrule(lr){4-6} \cmidrule(lr){7-9}
    & & & Pre. & Rec. & F1 & P-C & P-R & VP-S \\
    \midrule
    \textbf{Learning-Based} 
    & \textsc{DeepDFA} & \textsc{GNN-based} & 0.66 & 0.92 & 0.77 & 0.37 & 0.24 & 0.13 \\
    \midrule
    \multirow{2}{*}{\textbf{Agent-Based} }
    & \textsc{VulTrial} & \textsc{DeepSeek V3.2} & \todo{0.50} & \todo{0.49} & \todo{0.49} & \todo{0.17} & \todo{0.16} & \todo{0.01} \\
    & \textsc{JitVul} & \textsc{DeepSeek V3.2} & \revision{0.51} & \revision{0.85} & \revision{0.64} & \revision{0.12} & \revision{0.07} & \revision{0.05} \\
    \midrule
    \multirow{5}{*}{\textbf{LLM-Based}} 
    & \textsc{CoT} & \textsc{DeepSeek V3.2} & 0.59 & 0.64 & 0.61 & 0.16 & 0.08 & 0.08 \\
    & \textsc{LLMxCPG} & \textsc{Qwen2.5-32B} & 0.51 & \textbf{0.99} & 0.67 & 0.47 & 0.30 & 0.17 \\
    & \textsc{VulRAG} & \textsc{DeepSeek V3.2} & \todo{0.56} & \todo{0.53} & \todo{0.54} & \todo{0.24} & \todo{0.16} & \todo{0.08} \\
    & \textsc{VulInstruct} & \textsc{DeepSeek V3.2} & \todo{0.50} & \todo{0.96} & \todo{0.66} & \todo{0.03} & \todo{0.02} & \todo{0.01} \\
    & \textbf{\tool} & \textsc{DeepSeek V3.2} & \todo{\textbf{\revision{0.75}}} & \todo{\revision{0.80}} & \todo{\textbf{\revision{0.78}}} & \todo{\textbf{\revision{0.51}}} & \todo{\textbf{\revision{0.00}}} & \todo{\textbf{\revision{0.51}}} \\
    \bottomrule
  \end{tabular*}
\end{table*}

\textbf{Overall Results.} As shown in Table~\ref{tab:generality_comparison}, \tool achieved a precision of \todo{\revision{0.75}}, recall of \todo{\revision{0.80}}, and F1-score of \todo{\revision{0.78}} on the \textsc{PrimeVul} dataset. Compared to the learning-based approach, \textsc{DeepDFA}, \tool improves precision by \todo{\revision{0.09}} (\todo{\revision{14}}\%) while maintaining a superior F1-score, despite a moderate trade-off in recall.
For agent-based approaches, \revision{\tool outperforms the strongest agent-based baseline \textsc{JitVul} by \todo{\revision{0.14}} (\todo{\revision{22}}\%) in F1-score, and exceeds \textsc{VulTrial} by \todo{\revision{0.29}} (\todo{\revision{59}}\%). This gap highlights that generic agentic workflows without specialized security-aware enhancements struggle to generalize to diverse vulnerability patterns.}
Furthermore, against the best state-of-the-art LLM-based approach \textsc{LLMxCPG}, which demonstrates high recall but low precision, \tool attains a much better trade-off, improving precision by \todo{\revision{0.24}} (\todo{\revision{47}}\%) and F1-score by \todo{\revision{0.11}} (\todo{\revision{16}}\%) over it.
This indicates that \tool more accurately identifies true vulnerabilities and avoids excessive false positives, addressing the ``over-warning'' issue prevalent in current LLM-based solutions. 
On pairwise metrics, \tool achieves a VP-S score of \todo{\revision{0.51}}, outperforming the best baseline \textsc{LLMxCPG} (\todo{\revision{0.17}}) by \todo{\revision{0.34}} (\todo{\revision{200}}\%), demonstrating its strong pairwise discriminative ability in distinguishing textually similar vulnerable and patched code.


\subsection{Efficiency Evaluation (RQ5)}
\begin{table}[!t]
    \centering
    \fontsize{7.0pt}{6.0pt}\selectfont
    \caption{Results of Our Efficiency Evaluation}
    \renewcommand{\arraystretch}{1.2}
    \label{tab:efficiency_comparison}
    \vspace{-10pt}
    \begin{tabular}{lccccccccc}
      \toprule
      & \textsc{DeepDFA} & \textsc{CoT} & \textsc{LLMxCPG} & \textsc{VulInstruct} & \textsc{VulRAG} & \textsc{JitVul} & \textsc{IRIS} & \textsc{VulTrial} & \tool \\
      \midrule
      \textbf{Time (s)} & \textbf{79.85} & 137.29 & 136.78 & 181.37 & 127.69 & \revision{254.76} & 397.29 & \revision{234.51} & \todo{\revision{188.51}} \\
      \textbf{Token (\#)} & -- & \textbf{3,435} & -- & 7,665 & 4,659 & \revision{9,669} & \revision{79,827} & \revision{11,659} & \revision{60,357} \\
      \bottomrule
    \end{tabular}
  \end{table}
\textbf{RQ5 Setup.} \cyh{We measured the average time and tokens consumed to detect vulnerabilities for each sample in the \textsc{PrimeVul4J} dataset. We also provide a phase-level breakdown of \tool's cost.}

\textbf{Overall Results.} Table~\ref{tab:efficiency_comparison} details the efficiency analysis. \tool requires \todo{\revision{188.51}} seconds per detection, notably slower than the fastest baseline, \textsc{DeepDFA}, which took \todo{\revision{79.85}} seconds. This overhead primarily stems from the rigorous static analysis based on Joern to construct the initial UDG. \camera{In terms of token consumption, the LLM-based approaches naturally fall into two categories: (1)~\emph{lightweight methods} (\textsc{CoT}, \textsc{VulRAG}, \textsc{VulInstruct}) that issue single or few LLM queries with minimal context (3K--8K tokens), and (2)~\emph{context-intensive methods} (\textsc{JitVul}, \textsc{VulTrial}, \textsc{IRIS}, \tool) that provide extensive context or invoke the LLM multiple times (10K--80K tokens). \tool belongs to the second category by design, as it deliberately invests tokens in UDG enhancement and holistic context construction to achieve grounded reasoning. Within this category, \tool consumes \todo{\revision{60,357}} tokens per sample, which is \todo{24}\% lower than \textsc{IRIS} (\todo{\revision{79,827}} tokens), while achieving substantially stronger detection performance, confirming that \tool utilizes its token budget far more effectively. Relative to lightweight methods, \tool's higher token cost is the direct consequence of providing richer vulnerability context and issuing additional queries for graph enhancement, a trade-off that yields significant effectiveness improvements as shown in Table~\ref{tab:effectiveness_and_ablation_comparison}. Those baselines that require training (\textsc{DeepDFA}, \textsc{LLMxCPG}) incur no inference-time token cost but deliver weaker performance, especially on pairwise metrics. Overall, we consider this cost justified given the substantial gains in detection effectiveness.}

\begin{table}[!t]
  \footnotesize
  \centering
  \caption{Phase-level Breakdown of \tool's Efficiency}
  \vspace{-10pt}
  \label{tab:breakdown}
  \begin{tabular}{lcccc}
    \toprule
    \textbf{Phase} & \textbf{Time (s)} & \textbf{Time (\%)} & \textbf{Token (\#)} & \textbf{Token (\%)} \\
    \midrule
    Unified Dependency Graph Construction & \revision{159.35} & \revision{84.5\%} & \revision{46,174} & \revision{76.5\%}\\
    Holistic Vulnerability Context Extraction & \revision{28.06} & \revision{14.9\%} & 0 & 0.0\%\\
    Context-Aware LLM Reasoning & \revision{1.10} & \revision{0.6\%} & \revision{14,183} & \revision{23.5\%}\\
    \midrule
    \textbf{Total} & \textbf{\revision{188.51}} & \textbf{100\%} & \textbf{\revision{60,357}} & \textbf{100\%} \\
    \bottomrule
  \end{tabular}
\end{table}
\cyh{\textbf{Phase-level Breakdown.} Table~\ref{tab:breakdown} presents the phase-level cost breakdown of \tool. UDG Construction dominates both runtime (\revision{84.5\%}) and token usage (\revision{76.5\%}), as the LLM is strategically invoked to resolve ambiguous polymorphic and reflective call targets. Holistic Context Extraction incurs zero token cost via deterministic graph traversal, contributing a modest fraction (\revision{14.9\%}) of total runtime. Context-Aware LLM Reasoning consumes the remaining \revision{23.5\%} of tokens yet only \revision{0.6\%} of runtime, confirming that grounded reasoning is computationally lightweight once holistic context is established.}
\subsection{Usefulness Evaluation (RQ6)}

\textbf{RQ6 Setup.} To assess the practical usefulness of \tool, we selected \todo{9} widely used Java projects from public code repositories that (1) had at least one CVE reported in the past year and (2) received active updates within the most recent month. \camera{These projects span diverse application domains and we scanned the specific release tags listed in Table~\ref{tab:usefulness_comparison} for reproducibility.}
Details of these projects are provided in Table~\ref{tab:usefulness_comparison}. Due to the poor performance and high false positive rates of other tools on public~datasets, which would impose excessive manual validation costs, we used only \tool~to~detect~vulnerabilities in these projects, identifying \todo{38} candidates. These were manually verified by two security experts with over 5 years of experience, achieving a Cohen's Kappa agreement of \todo{0.934}, with a third expert resolving discrepancies. This yielded \todo{26} vulnerable samples and \todo{12} non-vulnerable~ones.

\textbf{Overall Results.} \cyh{On real-world open source projects, \tool achieved a precision of \todo{0.68}, significantly reducing manual verification effort. \revision{Equivalently, using FDR\,=\,FP/(FP$+$TP), the open-source scan yields an FDR of \todo{0.32} (\todo{12}/(\todo{12}$+$\todo{26}))}. Notably, all \todo{26} true positives are previously unknown vulnerabilities. We therefore responsibly disclosed them to the respective project maintainers by filing security issues, and \todo{15} have been confirmed, with others under review. Additionally, \todo{6} vulnerabilities were submitted for CVE assignment, resulting in~\todo{5}~CVE~identifiers.}

To further evaluate \tool in an industrial setting, we collaborated with a leading technology company serving over a billion users. \revision{The deployment followed an offline audit workflow, i.e., \tool scanned the internal repository, generated per-candidate reports containing the target function, extracted vulnerability context, and structured reasoning output, and security experts together with project developers triaged and confirmed the findings.} In a critical internal Java repository, \tool detected \todo{40} confirmed vulnerabilities \revision{with an FDR of \todo{0.17}}. Due to confidentiality agreements, details of these industrial cases cannot be disclosed.

\revision{\textbf{Human Study.} To quantify how \tool's output reduces manual triage effort, we conducted a human study with \todo{10} developers, each with at least two years of Java development experience. Every participant received the same \todo{10} vulnerability candidates randomly sampled from the detected \todo{26} vulnerable samples in the real world, evenly split into two fixed groups, i.e., 5 candidates triaged under Condition~(1) (w/o \tool), where only the target function and a vulnerability warning were provided, and the remaining 5 under Condition~(2) (w/ \tool), where each target function was augmented with \tool's structured reasoning output and the extracted vulnerability context. The presentation order of candidates was randomized across participants to mitigate ordering and practice effects. We measured the average triage time and the accuracy of each participant's verdict against the ground truth. Table~\ref{tab:human-study} reports the results. With \tool's augmented output, participants reduced their average triage time by \todo{59\%} and improved verdict accuracy by \todo{27\%}, demonstrating that the structured reasoning and context provided by \tool substantially lower the barrier to vulnerability confirmation.}

\begin{table}[!t]
    \centering
    \fontsize{7.0pt}{6.0pt}\selectfont
    \caption{\revision{Results of Our Human Study}}
    \renewcommand{\arraystretch}{1.2}
    \label{tab:human-study}
    \vspace{-10pt}
    \begin{tabular}{lcc}
      \toprule
      & \textbf{w/o \tool} & \textbf{w/ \tool} \\
      \midrule
      Avg. Time (s) & \todo{3,791} & \todo{1,543} \\
      Accuracy & \todo{0.66} & \todo{0.84} \\
      \bottomrule
    \end{tabular}
\end{table}

\revision{\textbf{Developer Feedback.} To assess the usability of \tool in practice, we surveyed \todo{5} developers from the collaborating company whose projects had been analyzed by \tool. All respondents confirmed that \tool improved their projects' security posture and expressed interest in incorporating it into their regular development workflow. In terms of improvement suggestions, they highlighted two directions, i.e., (1)~augmenting detection reports with reproducible Proof-of-Concept (PoC) exploits to accelerate vulnerability confirmation and remediation, and (2)~modularizing \tool's core capabilities (e.g., context extraction, and vulnerability reasoning) into agent-callable components, enabling seamless integration with emerging LLM agent frameworks for automated, end-to-end vulnerability detection and remediation.}
\begin{table}[!t]
    \centering
    \fontsize{7.0pt}{6.0pt}\selectfont
    \caption{Results of Our Usefulness Evaluation}
    \renewcommand{\arraystretch}{1.2}
    \label{tab:usefulness_comparison}
    \vspace{-10pt}
    \begin{tabular}{lccccc}
      \toprule
      \textbf{Project Name} & \textbf{Version} & \textbf{Lines of Code (\#)} & \textbf{Detected (\#)} & \textbf{True Positives (\#)} & \textbf{Confirmed (\#)} \\
      \midrule
      pf4j/pf4j & release-3.14.1 & 9.9k & 3 & 2 & 1 \\
      traccar/traccar & v6.11.1  & 91.0k & 4 & 3 & 2 \\
      codehaus-plexus/plexus-utils & plexus-utils-4.0.2 & 13.5k & 2 & 1 & 1 \\
      apache/flink & release-2.2.0 & 1,668.8k & 3 & 2 & 1 \\
      apache/dubbo & dubbo-3.3.6 & 307.9k & 2 & 1 & 1 \\
      apache/pulsar & v4.1.2 & 671.5k & 1 & 1 & 1 \\
      jeecgboot/JeecgBoot & v3.9.0 & 60.6k & 9 & 7 & 2 \\
      yangzongzhuan/RuoYi & v4.8.2 & 23.5k & 8 & 5 & 3 \\
      makunet/maku-boot & v4.11.0 & 15.6k & 6 & 4 & 3 \\
      \midrule
      Total &  & 2,862.3k & 38 & 26 & 15 \\
      \bottomrule
    \end{tabular}
  \end{table}

\subsection{UDG Enhancement Evaluation (RQ7)}
\cyh{\textbf{RQ7 Setup.} To evaluate the accuracy of UDG enhancement, we analyze the structural modifications across the \todo{9} selected Java projects in \textbf{RQ6}. Our approach resulted in a total of \todo{86,382} added edges, where the augmentation primarily focused on call edges (\todo{83,568}), control flow edges (\todo{29}), and data dependency edges (\todo{2,785}). Simultaneously, the refinement process identified and removed \todo{292,286} redundant or incorrect edges, specifically targeting \todo{83,123} call edges and \todo{209,163} data dependency edges. Due to the prohibitive cost of exhaustive verification for large-scale modifications, we employed stratified random sampling based on a \todo{95\%} confidence level and a \todo{5\%} margin of error for call and data dependency modifications. For the control flow edges, we performed an exhaustive verification of all \todo{29} added edges due to their small quantity. This strategy yielded \todo{749} added edges for validation (comprising \todo{382} call, \todo{29} control flow, and \todo{338} data dependency edges) alongside \todo{765} removed edges (\todo{382} call and \todo{383} data dependency edges). Each sampled edge was independently inspected by two security experts with over \todo{5} years of experience to verify its alignment with source code semantics. Any conflicting judgments were resolved by a third expert arbitrator, ultimately achieving an overall Cohen's Kappa coefficient of \todo{0.958}.}
\begin{table}[!t]
  \footnotesize
  \centering
  \caption{UDG Enhancement Accuracy by Edge Type}
  \vspace{-10pt}
  \label{tab:udg-accuracy}
  \begin{tabular}{lllcccc}
    \toprule
    \textbf{Edge Type} & \textbf{Strategy} & \textbf{Op.} & \textbf{Total (\#)} & \textbf{Sampled} & \textbf{Correct} & \textbf{Acc.} \\
    \midrule
    \multirow{2}{*}{Call} & \multirow{2}{*}{Neuro-Symbolic} & Added & \todo{83,568} & \todo{382} & \todo{322} & \todo{0.84} \\
    & & Removed & \todo{83,123} & \todo{382} & \todo{375} & \todo{0.98} \\
    Control Flow & Deterministic & Added & \todo{29} & \todo{29} & \todo{28} & \todo{0.97} \\
    \multirow{2}{*}{Data Dependency} & \multirow{2}{*}{Deterministic} & Added & \todo{2,785} & \todo{338} & \todo{327} & \todo{0.97} \\
    & & Removed & \todo{209,163} & \todo{383} & \todo{360} & \todo{0.94} \\
    \midrule
    \multirow{2}{*}{\textbf{Overall}} & \multirow{2}{*}{Neuro-Symbolic} & Added & \todo{86,382} & \todo{749} & \todo{677} & \todo{0.90} \\
    & & Removed & \todo{292,286} & \todo{765} & \todo{735} & \todo{0.96} \\
    \bottomrule
  \end{tabular}
\end{table}

\cyh{\textbf{Overall Results.} Table~\ref{tab:udg-accuracy} presents the accuracy of UDG enhancement broken down by edge type.}

\cyh{\emph{Call Edge Enhancement.}
Reflective calls are resolved to add edges, and infeasible polymorphic dispatches are pruned to remove them; expert validation reports accuracies of \todo{0.84} and \todo{0.98}, respectively. Addition is harder because opaque third-party code and incomplete type hierarchies force the LLM to guess concrete reflection targets, which sometimes yields unsupported edges. The same gaps can bias polymorphic resolution, but certifying an infeasible dispatch is typically easier than inferring a reflective target, consistent with the higher removal accuracy. Most errors arise in heavily library-driven, dynamic code sections where limited source visibility restricts both neural and traditional static analyses. The high pruning accuracy confirms that \tool effectively strips spurious call structure before downstream vulnerability reasoning.}

\cyh{\emph{Control Flow Edge Enhancement.}
\tool deterministically reconstructs missing control flow edges for labeled \texttt{break} and \texttt{continue} statements using Tree-sitter. Since this is a rule-based, syntax-directed procedure that precisely resolves jump targets from the AST, it achieves an accuracy of \todo{0.97}. The few errors are attributed to rare corner cases involving deeply nested labeled constructs with complex scoping rules.}

\cyh{\emph{Data Dependency Edge Enhancement.}
Data dependency enhancement involves two deterministic operations. For edge addition, \tool constructs new dependency edges for global variable assignments via Tree-sitter, achieving an accuracy of \todo{0.97}. The remaining errors primarily arise from implicit side-effects in static initializers (e.g., \texttt{static\{...\}} blocks), where incomplete type binding leads to missing dependency chains. For edge removal, the bottom-up, summary-based taint analysis prunes spurious inter-procedural dependencies introduced by Joern's conservative heuristics, achieving an accuracy of \todo{0.94}, which validates the soundness of function summary propagation via Tarjan's SCC-based ordering. The remaining errors mainly stem from annotation-driven code generation (e.g., Lombok), where implicitly synthesized methods are invisible to the analysis.}

\cyh{Overall, \tool achieves a high accuracy of \todo{0.90} for edge addition and \todo{0.96} for edge removal, with a pooled validation accuracy of approximately \todo{0.93} over all sampled modifications, demonstrating that the neuro-symbolic enhancement strategy effectively improves UDG quality and provides a reliable foundation for vulnerability detection.}
\camera{We clarify that this accuracy measures sampled edge-level semantic correctness rather than a formal soundness guarantee for downstream detection. The confidence threshold $\theta_c$ and the ablation study (w/o $\mathcal{G}_e$) confirm that UDG enhancement consistently improves detection performance despite potential incorrect edges.}


\section{Discussion}\label{sec:discussion}
\revision{\textbf{Internal Validity.} The first concern is the non-deterministic nature of LLM inference. \tool mitigates this by aggregating $\mathcal{N}$ independent predictions via majority voting and filtering enhancement decisions below a confidence threshold $\theta_c$. Our sensitivity analysis confirms that performance stabilizes at $\mathcal{N}=3$ and $\theta_c=8$.
The second concern is potential data leakage, as the LLM's training corpus may contain known CVE information. We address this via a strict chronological split in \textsc{PrimeVul4J}, ensuring all test samples postdate the training set. 
The third concern is baseline fairness. \tool enhances the UDG produced by Joern, whereas some baselines consume Joern's raw output directly. This difference is by design, as UDG enhancement is a core contribution of \tool. The ablation w/o $\mathcal{G}_e$ isolates this effect, and \tool still outperforms baselines that do not depend on Joern at all.}

\revision{\textbf{External Validity.} The primary threat concerns dataset scale and representativeness. We followed the rigorous \textsc{PrimeVul} methodology~\cite{ding2024vulnerability} to ensure high label quality, which resulted in a relatively compact test set of \todo{168} samples. Nevertheless, \tool maintained consistent performance (F1\,=\,0.75, VP-S\,=\,0.54) across the entire \textsc{PrimeVul4J} dataset, and our additional evaluations on the C/C++ \textsc{PrimeVul} dataset and \todo{9} real-world Java projects further mitigate dataset-specific bias.
A secondary threat stems from the commit-based labeling methodology, where pre-fix code is labeled as vulnerable and post-fix code as non-vulnerable. This may be affected by tangled commits or incomplete fixes. We mitigate this through the PrimeVul denoising pipeline, which filters out ambiguous samples. Additionally, all vulnerability candidates in the usefulness evaluation were independently verified by two security experts (Cohen's Kappa\,=\,\todo{0.934}), providing orthogonal validation of label reliability.}

\revision{\textbf{Limitations and Future Work.}}
\revision{Despite the promising performance of \tool, four limitations need to be~acknowledged.}
\revision{First, although the neuro-symbolic UDG enhancement framework is language-agnostic in design, its current heuristics are tuned for Java-specific features (e.g., reflection), and do not yet cover language-dependent patterns such as dynamic dispatch in Python/JavaScript. Our cross-language evaluation on C/C++ demonstrates the transferability of the overall framework, but does not imply full support for all language-specific features. As future work, we intend to develop dispatch resolution strategies for additional languages that reuse the core neuro-symbolic inference while adapting candidate identification to each language's semantics.}

\revision{Second, the CFG enhancement currently targets labeled jump constructs (e.g., \texttt{break}/\texttt{continue} with labels and \texttt{goto}). Other control flow such as exception handling and try-with-resources is already modeled by the underlying static analysis tool. However, asynchronous callbacks (e.g., \texttt{CompletableFuture}, and event listeners) introduce implicit cross-function control flow that cannot yet be precisely captured. As future work, we plan to incorporate event-driven control flow modeling and leverage LLM agents to explore callback chains on demand.}

\revision{Third, \tool currently relies on a static slicing strategy with predefined sensitive APIs, which may include extraneous context that triggers the ``lost-in-the-middle'' phenomenon and cannot yet automatically discover new sinks. As future work, potential directions to mitigate this include importance-aware context ranking that prioritizes critical fragments (e.g., defense logic and source-sink paths), hierarchical summarization that condenses peripheral context before feeding it to the LLM, and staged reasoning that decomposes long contexts into focused sub-queries. We also aim to modularize the core capabilities into agent-callable skills, leveraging LLM agents for adaptive, on-demand context extraction and automated sink discovery.}

\camera{Finally, meta-prompt construction depends on an expert-curated CWE knowledge base currently covering \todo{102} primary CWE types. The tool's precision is constrained by the coverage and granularity of these knowledge sources. In practice, adding a new CWE type requires approximately one hour of expert effort to define the detection reasoning steps and sensitive API mappings. Since the CWE guidelines prescribe abstract reasoning steps (e.g., taint propagation tracing and defense adequacy assessment) rather than concrete API names, they remain stable across framework and library updates. Only the sensitive API mappings require periodic maintenance when new frameworks are adopted. As future work, we plan to expand the knowledge base to long-tail vulnerability types and develop automated knowledge extraction and prompt generation methods.}


\section{Related Work}\label{sec:related-work} 
\textbf{Learning-based Vulnerability Detection.} Learning-based vulnerability detection has evolved from sequential to structural modeling. Early approaches relied on LSTM and BiLSTM to extract vulnerability patterns from code slices \cite{li2016vulpecker,li2018vuldeepecker,li2021sysevr}, capturing local semantics but missing complex code structures. Graph Neural Networks (GNNs) advanced the field by representing code as graphs (e.g., AST, CFG, PDG) to better capture structural dependencies \cite{zhou2019devign,cheng2021deepwukong,steenhoek2024dataflow}. Nonetheless, both sequence-based and GNN-based approaches struggle with inter-procedural vulnerabilities. To address this, \textsc{SnapVul} \cite{wu2023learning} introduced vulnerability type specific inter-procedural slicing, while \textsc{VulnSC} \cite{wu2025enhancing} leveraged LLM-generated function summaries to enrich context.
However, both approaches remain limited by their reliance on static analysis representations, which can result in fragmented and incomplete context. Slicing strategies may also overlook implicit semantic dependencies, leading to missed vulnerability contexts.

\textbf{Large Language Models for Vulnerability Detection.}
Large Language Models (LLMs) have introduced new paradigms for vulnerability detection through approaches such as prompting \cite{nong2024chain, steenhoek2024comprehensive, tamberg2025harnessing, ullah2024llms, zhang2024prompt, zhou2024large}, in-context learning \cite{lu2024grace}, and fine-tuning \cite{du2024generalization, yang2024security}. 
Despite promising performance, these approaches often struggle to differentiate vulnerable code from its patched counterparts. Ding et al.~\cite{ding2024vulnerability} introduced the \textsc{PrimeVul} dataset and showed that even state-of-the-art LLMs mainly exploit superficial patterns rather than capturing true semantic differences.

To mitigate these issues, recent work has sought to enrich vulnerability detection with broader context. Du et al.~\cite{du2024vul} incorporate historical vulnerability information, while Zhu et al.~\cite{zhu2025specification} mine security specifications from such history. \revision{Li et al.~\cite{li2025no} combine lightweight function-level slicing with RAG-enhanced multi-layer LLM reasoning for smart contract auditing.} Widyasari et al.~\cite{widyasari2025let} introduce a multi-agent, courtroom-inspired framework with specialized roles. \revision{Wang et al.~\cite{wang2026security} propose a training-free multi-agent framework that transforms vulnerability detection into contract verification by reverse-engineering Gherkin behavioral specifications from vulnerability-fix pairs.} Despite these advances, existing approaches generally lack explicit step-by-step reasoning mechanisms and self reflection for LLMs, leading to inaccurate results and limited improvements on \textsc{PrimeVul}~\cite{zhu2025specification}. 
\cyh{Moreover, most LLM-based approaches consider only a single-function context, neglecting inter-procedural information. Some recent works address this by leveraging static analysis tools (e.g., Joern~\cite{joern}, CodeQL~\cite{codeql}) to provide inter-procedural context~\cite{lekssays2025llmxcpg, li2024iris, guo2025repoaudit}. However, their reliance on static representations often yields context that is either over-approximated or disconnected from runtime semantics (e.g., missing data flow in complex references), which limits the LLM's ability to reason about actual exploitability. 
Although Wang et al.~\cite{wang2024llmdfa} combine LLMs with data-flow analysis to better trace inter-procedural semantic dependencies, their method primarily models data-flow signals while underrepresenting fragmented control-flow information. Moreover, it supports only three bug types and relies heavily on manually crafted rules.}

\revision{In contrast to these prior efforts, which directly use the program representation produced by static analysis tools without further refinement, \tool integrates LLM-based semantic inference into the construction of the program representation itself. Specifically, \tool leverages LLMs to enhance the accuracy of the unified dependency graph (e.g., resolving polymorphic and reflective dispatch), and then extracts holistic vulnerability context from this enhanced graph for detection. Combined with vulnerability type specific meta-prompting that enforces structured, evidence-driven reasoning rather than generic CoT, this design allows \tool to simultaneously mitigate static analysis inaccuracies and ground LLM reasoning in precise program semantics.}

\section{Conclusions}\label{sec:conclusion}
We present \tool, an LLM-based approach that weaves broken program semantics into accurate representations for grounded vulnerability detection. \tool constructs an enhanced Unified Dependency Graph (UDG) for a given repository using neuro-symbolic techniques, extracts holistic vulnerability context including explicit and implicit vulnerability context, and employs meta-prompting with self-consistency reasoning to enable accurate LLM-based detection. Extensive experiments confirm \tool's effectiveness, generality, and practical usefulness. 

\section{Data Availability}

The source code for \tool, with experimental data and results, is available at our website~\cite{opensource}.

\section*{Acknowledgment}

This work was supported by the National Natural Science Foundation of China (Grant No. 62332005 and 62372114).

\bibliographystyle{ACM-Reference-Format}
\bibliography{src/references}

\end{document}